\documentclass[a4paper,fleqn,usenatbib]{mnras}

\usepackage{newtxtext,newtxmath}

\usepackage[T1]{fontenc}
\usepackage{ae,aecompl}

\usepackage{graphicx}	
\usepackage{amsmath}	
\usepackage{amssymb}	

\usepackage{ulem}
\usepackage{color}

\newcommand{\sse}{\texttt{SSE}}
\newcommand{\bse}{\texttt{BSE}}
\newcommand{\nbodyfour}{\texttt{NBODY4}}
\newcommand{\nbodysix}{\texttt{NBODY6}}

\newcommand{\teff}{\log(T_{\rm eff}/\mbox{K})}
\newcommand{\linz}{Z/Z_\odot}
\newcommand{\logz}{\log(Z/Z_\odot)}
\newcommand{\logr}{\log(R/R_\odot)}

\newcommand{\mhgu}{M_{\rm HG,u}}
\newcommand{\mblu}{M_{\rm BL,u}}
\newcommand{\mebu}{M_{\rm EB,u}}
\newcommand{\mebl}{M_{\rm EB,l}}
\newcommand{\mcbu}{M_{\rm CB,u}}

\title[Fitting formulae for evolution tracks of EMP stars]{Fitting
  formulae for evolution tracks of massive stars under extreme metal
  poor environments for population synthesis calculations and star
  cluster simulations}

\author[Ataru Tanikawa]{
A. Tanikawa,$^{1,2}$\thanks{E-mail: tanikawa@ea.c.u-tokyo.ac.jp}
T. Yoshida,$^{3}$
T. Kinugawa,$^{3,4}$
K. Takahashi,$^{5}$
H. Umeda$^{3}$
\\
$^{1}$Department of Earth Science and Astronomy, College of
  Arts and Sciences, The University of Tokyo, 3-8-1 Komaba, Meguro-ku,
  Tokyo 153-8902, Japan\\
$^{2}$RIKEN Advanced Institute for Computational Science,
  7-1-26 Minatojima-minami-machi, Chuo-ku, Kobe, Hyogo 650-0047,
  Japan\\
$^{3}$ Department of Astronomy, Graduate School of Science, The
  University of Tokyo, 7-3-1 Hongo, Bunkyo-ku, Tokyo 113-0033, Japan\\
$^{4}$ Institute for Cosmic Ray Research, The University of Tokyo,
  5-1-5 Kashiwa-no-ha, Kashiwa City, Chiba 277-8582, Japan \\
$^{5}$ Max Planck Institute for Gravitational Physics (Albert Einstein
  Institute), Am \"hlenberg 1, Postsdam-Golm 14476, Germany}

\date{Accepted XXX. Received YYY; in original form ZZZ}

\pubyear{2017}

\begin{document}
\label{firstpage}
\pagerange{\pageref{firstpage}--\pageref{lastpage}}
\maketitle

\begin{abstract}

We have devised fitting formulae for evolution tracks of massive stars
with $8 \lesssim M/M_\odot \lesssim 160$ under extreme metal poor
(EMP) environments for $\logz = -2, -4, -5, -6$, and $-8$, where
$M_\odot$ and $Z_\odot$ are the solar mass and metallicity,
respectively. Our fitting formulae are based on reference stellar
models which we have newly obtained by simulating the time evolutions
of EMP stars. Our fitting formulae take into account stars ending with
blue supergiant (BSG) stars, and stars skipping Hertzsprung gap (HG)
phases and blue loops, which are characteristics of massive EMP
stars. In our fitting formulae, stars may remain BSG stars when they
finish their core Helium burning (CHeB) phases. Our fitting formulae
are in good agreement with our stellar evolution models. We can use
these fitting formulae on the \sse, \bse, \nbodyfour, and
\nbodysix\;codes, which are widely used for population synthesis
calculations and star cluster simulations. These fitting formulae
should be useful to make theoretical templates of binary black holes
formed under EMP environments.

\end{abstract}

\begin{keywords}
gravitational waves -- binaries: general
\end{keywords}

\section{Introduction}
\label{sec:introduction}

Laser Interferometer Gravitational-Wave Observatory (LIGO) has finally
detected the first gravitational wave from a black hole (BH) merger
\citep{2016PhRvL.116f1102A}. Since then, many BH-BH mergers have been
observed by gravitational wave observatories LIGO and VIRGO
\cite[e.g.][]{2019PhRvX...9c1040A}. These detections have raised an
important question: what the origin of these merging BH-BHs is. One of
the promising origins is massive binary stars. However, it has been
still under debate what stellar metallicity such massive binary stars
have: Population (Pop.) I/II stars \cite[e.g.][]{2016Natur.534..512B}
or Pop.~III stars \cite[e.g.][]{2014MNRAS.442.2963K}, and where they
are formed: galactic fields
\cite[e.g.][]{1973NInfo..27....3T,1998ApJ...506..780B} or star
clusters \cite[e.g.][]{2000ApJ...528L..17P}. In order to elucidate the
origin of these merging BH-BHs, one has to make theoretical templates
of merging BH-BHs, and compare them with observed BH-BH populations.

Population synthesis calculations and star cluster simulations are
powerful tools to make such theoretical templates of merging BH-BHs
from galactic fields and from star clusters, respectively. In either
case, BH-BHs from Pop.~I/II stars with $0.01 \lesssim \linz \lesssim
1$ have been intensively studied so far
\cite[e.g.][]{2016Natur.534..512B,2016PhRvD..93h4029R}, where $Z$ and
$Z_\odot$ are metallicity and the solar metallicity, respectively. On
the other hand, BH-BHs formed from extreme metal poor (EMP) stars with
$\linz \lesssim 0.01$ (including Pop.~III stars) have been not
examined well.

\cite{2014MNRAS.442.2963K} have found that Pop.~III BH-BHs have
distinct features from Pop.~I/II BH-BHs by means of population
synthesis calculations. The mass distribution of Pop.~III BH-BHs have
a much larger peak than those of Pop.~I/II BH-BHs. This argument is
insensitive to the choices of stellar initial mass functions (IMFs)
and initial binary parameters \citep{2016MNRAS.456.1093K}.  Thus,
Pop.~III BH-BHs can have significant contribution to observed BH-BHs.
\cite{2017MNRAS.468.5020I} have confirmed their arguments by
simulating Pop.~III star evolutions. The reason for this difference
comes from stability of mass transfer of BH-BH progenitors. Massive
Pop.~I/II stars become red supergiant (RSG) stars, and have convective
envelopes after a certain time in their lives. Such stars easily
experience unstable mass transfer or common envelope evolution
\citep{1976IAUS...73...75P,1993PASP..105.1373I,2000ARA&A..38..113T,2013A&ARv..21...59I},
just after they begin Roche-lobe overflow. In fact, most of BH-BH
progenitors experience common envelope evolution for Pop.~I/II stars
\cite[e.g.][]{1998ApJ...506..780B,2002ApJ...572..407B,2012ApJ...759...52D,2013ApJ...779...72D,2014A&A...564A.134M,2014ApJ...789..120B,2015MNRAS.451.4086S,2016MNRAS.462.3302E,2016Natur.534..512B,2017PASA...34...58E,2017MNRAS.472.2422M,2017NatCo...814906S,2018MNRAS.479.4391M,2018MNRAS.480.2011G,2018MNRAS.481.1908K,2019MNRAS.485..889S,2019MNRAS.487....2M,2019MNRAS.482..870E}. On
the other hand, a significant fraction of massive Pop.~III stars end
with blue supergiant (BSG) stars which have radiative envelopes, since
they have small opacities
\cite[e.g.][]{2001A&A...371..152M,2008A&A...489..685E}. They tend to
undergo stable mass transfer when they interact with their companion
stars. Such stable mass transfer loses less stellar masses than common
envelope evolution for the following reason. If neither of two stars
are white dwarfs, neutron stars (NSs), and BHs, their total mass is
nearly conservative in stable mass transfer, while they can lose all
their envelopes in common envelope evolution as back reaction of the
tightening of the binary orbit. Hence, Pop.~III BH-BHs can be more
massive than Pop.~I/II BH-BHs.

Moreover, Pop.~III stars should have a different formation mode from
Pop.~I/II stars. Pop.~I/II stars have typical mass of $\sim 1M_\odot$
at the formation time, and top-light IMFs
\citep{1955ApJ...121..161S,2001MNRAS.322..231K}. On the other hand,
the typical mass of Pop.~III stars should be $10$ -- $1000M_\odot$ at
the initial time, and their IMF should be top-heavy
\citep{1998ApJ...508..141O,2002Sci...295...93A,2004ARA&A..42...79B,2008Sci...321..669Y,2011Sci...334.1250H,2011MNRAS.413..543S,2012MNRAS.422..290S,2013RPPh...76k2901B,2013ApJ...773..185S,2014ApJ...792...32S,2015MNRAS.448..568H}.
The formation mode may transition from Pop.~I/II like to Pop.~III like
at $Z/Z_\odot \sim 10^{-3}$ - $10^{-6}$
\citep{2003Natur.425..812B,2005ApJ...626..627O,2006MNRAS.369..825S,2010MNRAS.407.1003M}.
IMFs will be an important factor to amplify the difference between
Pop.~I/II and Pop.~III BH-BHs. We again emphasize that
the typical masses of Pop.~III BH-BHs in
  \cite{2014MNRAS.442.2963K} are mostly unchanged even if the
  top-heavy IMF is changed to Pop.~I/II IMFs.

Since Pop.~III BH-BHs have distinct features from Pop.~I/II BH-BHs, it
is instructive to bridge the metallicity gap between Pop.~I/II and
Pop.~III stars, and make templates of BH-BHs originating from EMP
stars with $0 \lesssim Z/Z_\odot \lesssim 0.01$. Such templates can
constrain the dominant metal environments under which BH-BH
progenitors are formed. Even if Pop.~III and EMP environments are not
dominant \citep{2016MNRAS.460L..74H,2017MNRAS.471.4702B}, such
templates will help surveying Pop.~III BH-BHs from an enormous number
of merging BH-BHs in current and future gravitational wave
observations \citep{2016PTEP.2016i3E01N}. The direct detection of
Pop.~III stars and their remnants have neither yet succeeded for
massive and short-lived Pop.~III stars \citep{2013MNRAS.429.3658R},
nor for low-mass and long-lived Pop.~III stars
\citep{2015ARA&A..53..631F}, although the latter Pop.~III stars might
be observed as metal-enriched stars due to metal pollution by
interstellar gas, dust, and asteroids
\citep{2015ApJ...808L..47K,2015MNRAS.453.2771J,2018PASJ...70...80T,2019MNRAS.486.5917K}.

In this paper, we devise evolution tracks of massive EMP stars for
population synthesis calculations and star cluster simulations, based
on stellar evolution simulations for massive stars with $8 \le
M/M_\odot \le 160$, where $M$ and $M_\odot$ are stellar mass and the
solar mass. There are many evolution tracks: \sse\;and
\bse\;\citep{2000MNRAS.315..543H,2002MNRAS.329..897H}, \texttt{SeBa}
\citep{1996A&A...309..179P}, Scenario Machine
\citep{1996A&A...310..489L}, \texttt{StarTrack}
\citep{2002ApJ...572..407B}, and \texttt{BINARY\_C},
\citep{2018MNRAS.473.2984I} driven by fitting formulae, and
\texttt{SEVN} \citep{2015MNRAS.451.4086S,2019MNRAS.485..889S},
\texttt{BPASS} \citep{2016MNRAS.462.3302E}, and \texttt{COMBINE}
\citep{2018MNRAS.481.1908K} based on detailed evolutionary
models. However, these evolution tracks support Pop.~I/II stars with
$0.001 \lesssim \linz \lesssim 1$. \cite{2014MNRAS.442.2963K} have
supported evolution tracks of just Pop.~III stars (i.e. $\linz = 0$),
based on the model of \cite{2001A&A...371..152M}. Our evolution tracks
support EMP stars with $\linz = 10^{-2}, 10^{-4}, 10^{-5}, 10^{-6}$,
and $10^{-8}$, and bridge the metallicity gap. For EMP stars, the
evolution tracks of $Z=10^{-4}, 10^{-6}, 10^{-10}$ and $0.7 \le M \le
15 M_\odot$ stars have been investigated \citep{1993ApJS...88..509C}.
The metallicity dependence of 20 $M_\odot$ stars with $Z=10^{-8},
10^{-5}, 0.02$ have been investigated in
\cite{2007A&A...461..571H}. However, no systematic studies of the
evolution tracks for EMP massive stars have been performed.

We preferentially make evolution tracks of massive EMP stars, since
stars should be dominantly formed as massive stars in EMP
environments. However, we will make evolution tracks of low-mass EMP
stars in near future. Many studies have claimed that low-mass stars
could be formed even under metal-free environments
\citep{2001ApJ...548...19N,2008ApJ...677..813M,2011ApJ...727..110C,2011Sci...331.1040C,2011ApJ...737...75G,2012MNRAS.424..399G,2013MNRAS.435.3283M,2014ApJ...792...32S,2016MNRAS.463.2781C,2019ApJ...877...99S}.

We have developed our evolution tracks as forms of fitting formulae in
order to incorporate the evolution tracks into \sse\;
\citep{2000MNRAS.315..543H}, \bse\;\citep{2002MNRAS.329..897H}, and
\nbodyfour\;and
\nbodysix\;\citep{2003gnbs.book.....A,2012MNRAS.424..545N,2015MNRAS.450.4070W}. Several
population synthesis calculation codes
\cite[e.g.][]{2002ApJ...572..407B,2014MNRAS.442.2963K,2018MNRAS.480.2011G}
are based on the \bse\;code. The \nbodyfour\;and \nbodysix\;codes are
widely used to derive BH-BH populations originating from star clusters
\cite[e.g.][]{2010MNRAS.402..371B,2013MNRAS.435.1358T,2014MNRAS.440.2714B,2017MNRAS.467..524B,2017PASJ...69...94F,2017MNRAS.469.4665P,2018MNRAS.480.5645H,2019MNRAS.486.3942K,2019MNRAS.487.2947D,2019arXiv191101434D}. Moreover,
many works have obtained BH-BH populations formed in star clusters,
using star cluster simulation codes coupled with the \bse\;code
\cite[e.g.][]{2013MNRAS.431.2184G,2018ComAC...5....5R}. Therefore, we
believe that our evolution tracks can be used on many codes for
population synthesis calculations and star cluster simulations with
minor adjustments.

The structure of this paper is as follows. In
section~\ref{sec:stellarEvolutionModels}, we overview our evolution
models of EMP stars as reference models of our evolution tracks. In
section~\ref{sec:implementation}, we describe how to make the fitting
formulae for the evolution tracks of EMP stars. In
section~\ref{sec:demonstration}, we compare our fitting formulae with
our stellar evolution models. In section~\ref{sec:summary}, we
summarize this paper. The units of time, luminosity, radius, and mass
are Myr, $L_\odot$ (the solar luminosity), $R_\odot$ (the solar
radius), and $M_\odot$, respectively, if otherwise specified.

\section{Stellar evolution models}
\label{sec:stellarEvolutionModels}

We need stellar evolution models as reference, in order to make
fitting formulae. We present our simulation method to make the stellar
evolution models in section~\ref{sec:SimulationMethod}. In
section~\ref{sec:SimulationResults}, we overview our stellar evolution
models.

\subsection{Simulation method}
\label{sec:SimulationMethod}

Our simulation method is similar to 1 dimensional (1D) simulation
method in \citet{Yoshida19}. We follow the time evolutions of stars
with $M=$ 8, 10, 13, 16, 20, 25, 32, 40, 50, 65, 80, 100, 125, and 160
$M_\odot$ for $\log (Z/Z_\odot)$ = -2, -4, -5, -6 and $-8$ from the
zero age main-sequence (ZAMS) to the carbon ignitions at the stellar
centers. We use a 1D stellar evolution code, {\tt HOSHI} code
\citep{2016MNRAS.456.1320T,2018ApJ...857..111T,Takahashi19,Yoshida19}.
Here we describe some details for the chemical mixing by convection
and input physics in the stellar evolution model. We do not account
for rotation and rotational mixing.

We adopt the Ledoux criterion for convective instability, and model
chemical mixing in a convective region by means of the mixing length
theory with diffusion coefficients as described in
\citet{Takahashi19}.  The diffusion coefficient in the convective
region is described as
\begin{equation}
D_{\rm cv} = \frac{1}{3} v_{\rm mix} l_{\rm mix},
\end{equation}
where $v_{\rm mix}$ is the velocity of convective blobs, $l_{\rm mix}
= \alpha_{\rm mix} H_P$ is the mixing length, and $H_P$ is the
pressure scale height.  The velocity of convective blobs is determined
by the mixing length theory \citep{Boehm58}.  The mixing length
parameter $\alpha_{\rm mix}$ is set to be 1.8.  In semiconvection
region, we model diffusive chemical mixing with the diffusion
coefficient derived in \citet{Spruit92},
\begin{equation}
D_{\rm sc} = f_{\rm sc} \frac{\nabla_{\rm rad} - \nabla_{\rm
    ad}}{(\varphi/\delta) \nabla_{\mu}} D_{{\rm therm}},
\end{equation}
where $f_{\rm sc}$ is a parameter corresponding to the square root of
the ratio of solute diffusivity to thermal diffusivity, $\nabla_{\rm
  rad} \equiv (\kappa L / 16\pi c GM_r) (3 P/a T^4)$ is a radiative
gradient, $\kappa$ is opacity, $L$ is luminosity, $c$ is the speed of
light, $G$ is the gravitational constant, $M_r$ is the mass
coordinate, $P$ is pressure, $a$ is the radiation constant, $T$ is
temperature, $\nabla_{\rm ad} \equiv (\partial \ln T/\partial \ln
P)_{s,\mu}$ is an adiabatic gradient, $s$ is entropy, $\mu$ is mean
molecular weight, $\nabla_{\mu} \equiv d \ln \mu / d \ln P$ is a
spatial gradient of mean molecular weight and $P$ is taken as a
measure of depth, $\varphi \equiv (\partial \ln \rho/\partial \ln
\mu)_{P,T}$ and $\delta \equiv - (\partial \ln \rho / \partial \ln
T)_{P,\mu}$ are thermodynamic derivatives, $\rho$ is density, thus
$\varphi/\delta = (\partial \ln T / \partial \ln \mu)_{P, \rho}$,
$D_{\rm therm} = 4 a c T^3 / 3 \kappa C_P$ is the thermal diffusivity,
$C_P$ is the specific heat by unit mass at constant pressure
\citep[e.g.][]{1990sse..book.....K,2009pfer.book.....M}.  The
parameter $f_{\rm sc}$ is set to be 0.3, which is the same value in
\citet{Takahashi19} \citep[see also][]{1999ApJ...513..861U, Umeda08}.

We also take into account chemical mixing by convective overshoot as a
diffusive process above convective regions. The diffusion coefficient
of the overshoot chemical mixing exponentially decreases with the
distance from the convective boundary as
\begin{equation}
D^{\rm ov}_{\rm cv} = D_{\rm cv,0} \exp \left( -2 \frac{\Delta
  r}{f_{\rm ov} H_{P0}} \right) ,
\end{equation}
where $D_{\rm cv,0}$ and $H_{P0}$ are the diffusion coefficient and
the pressure scale height at the convective boundary, respectively,
and $\Delta r$ is the distance from the boundary.  The overshoot
parameter $f_{\rm ov}$ is set to be 0.03, which is the same as the
value of Set L$_{\rm A}$ in \citet{Yoshida19}.  This overshoot
parameter is determined based on the calculation to early-B type stars
in the Large Magellanic Cloud similarly to Stern model
\citep{2011A&A...530A.115B}. The main-sequence width of a
solar-metallicity 20 $M_\odot$ model is almost same as that of the
corresponding Stern model \citep[see Fig. 12 ($a$) in][]{Yoshida19}.

For nuclear reaction network, 49 species of nuclei are taken into
account throughout evolution calculations\footnote{The adopted nuclear
  species are as follows: $^1$n, $^{1-3}$H, $^{3,4}$He, $^{6,7}$Li,
  $^{7,9}$Be, $^{8,10,11}$B, $^{11-13}$C, $^{13-15}$N, $^{15-18}$O,
  $^{17-19}$F, $^{20}$Ne, $^{23}$Na, $^{24}$Mg, $^{27}$Al, $^{28}$Si,
  $^{31}$P, $^{32}$S, $^{35}$Cl, $^{36}$Ar, $^{39}$K, $^{40}$Ca,
  $^{43}$Sc, $^{44}$Ti, $^{47}$V, $^{48}$Cr, $^{51}$Mn, $^{52-56}$Fe,
  $^{55,56}$Co, $^{56}$Ni.}.  We include thermonuclear reactions and
weak interactions ($\beta$ decays and electron captures) concerning to
the nuclear species, so the evolution calculation from hydrogen
burning (pp-chain and CNO cycles) until core-collapse (Si burning and
photo-disintegration) is available.  The rates of thermonuclear
reactions are adopted from JINA REACLIB v1 \citep{2010ApJS..189..240C}
except for $^{12}$C($\alpha, \gamma)^{16}$C.  The rate of
$^{12}$C($\alpha, \gamma)^{16}$O is adopted from \citet{Caughlan88}
and is multiplied by a factor of 1.2 \citep{2018ApJ...857..111T}.

We derive the initial composition of massive stars with a given
metallicity from the mixture of the solar-system composition and the
primordial chemical composition.  The elemental composition of the
solar system is evaluated from the bulk composition of the Sun in
\citet{2009ARA&A..47..481A}.  The mass fractions of hydrogen, helium,
and heavier elements are $X=0.7155$, $Y=0.2704$, and $Z=0.0141$,
respectively.  The isotopic ratio of each element is taken from the
table of the solar-system abundance derived from meteoritic and solar
photosphere data in \citet{Lodders09}.  The primordial chemical
composition is adopted from Eqs. (22)--(25) in \citet{Steigman07}.
EMP stars do not necessarily have chemical abundance scaled down from
the solar-system abundance \citep[see][for
  review]{2015ARA&A..53..631F}. However, there is no conclusive
abundance pattern for EMP stars. Thus, we adopt the scaling down of
the solar-system abundance conservatively.

We use the equation of state of \citet{Blinnikov96}, which includes
the effects of electrons, positrons, ions, and photons.  In low
temperature and density region, we use a non-degenerate equation of
state for electrons, ions, atoms, molecules, and photons taking
account of partial ionization and molecular dissociation
\citep{Vardya60, Iben63}.  We evaluate opacity using tables of OPAL
opacity \citep{Iglesias96}, molecular opacity \citep{Ferguson05}, and
conductive opacity \citep{Cassisi07}.  We evaluate the energy loss by
neutrinos using approximation formula in \citet{Itoh96}.

We have to extract data of He and CO core masses from the simulation
results. For this purpose, we define the He and CO cores of stars as
follows. The He-core mass is determined as the mass coordinates at the
outermost region where the hydrogen mass fraction is less than
0.1. The CO-core mass is determined as the mass coordinates at the
outermost region where the helium mass fraction is less than 0.1.

We do not include stellar wind mass loss in our simulations, since we
usually consider stellar wind mass loss while following fitting
formulae in population synthesis calculations and star cluster
simulations \citep[e.g.][]{2000MNRAS.315..543H}.

We stop following stellar evolutions when carbon is ignited at the
stellar centers. We can take into account the post-carbon burning
evolutions by a post-processing way in population synthesis
calculations and star cluster simulations. This is because stars
evolve on small timescale after the carbon ignition.  We confirmed
that the expansion of the radius from the carbon burning until the
onset of the core collapse (when the central temperature reaches
$10^{9.9}$ K) is less than 2 \% in a BSG star of $M/M_\odot =13
M_\odot$ and $\log (Z/Z_\odot) = -8$.  As a demonstration, we
implement the post-carbon burning evolutions, such as effects of
core-collapse supernova (SN) explosion, pulsational pair instability
(PPI) before core-collapse SN explosion, and pair instability (PI)
SNe, in section~\ref{sec:RemnantPhase}.

Our stellar models have masses of $\le 160M_\odot$, smaller than
possible Pop.~III masses ($\gtrsim 200M_\odot$). Nevertheless, this
mass range should be sufficient for our purpose for the following
reason. Pop.~III stars are formed in protostellar clouds with $\sim
1000M_\odot$ \citep[e.g.][]{2003ApJ...592..645Y}. The protostellar
clouds are gravitationally fragmented, and a large part of the
protostellar clouds are ejected through star forming processes,
according to recent numerical simulations \citep[see][for a
  review]{2018PhR...780....1D}. Therefore, Pop.~III stars with
$\gtrsim 200M_\odot$ should not be typical, and should be single if
present. The evolutions of single massive Pop.~III stars must be
interesting, however stellar models with masses of $\le 160M_\odot$
should be enough to investigate BH-BH merger events currently
observed.

\subsection{Simulation results}
\label{sec:SimulationResults}

We briefly review the evolutions of massive Pop.~I/II stars including
stars with $\logz = -2$ before we see the simulation results. A star
starts from a main-sequence (MS) phase in which hydrogen is burned at
the center of the star. The beginning time of the MS is called the
ZAMS time. When hydrogen is burned out at the center, a helium (He)
core has been formed inside of the star. Then, the He core and its
hydrogen envelope shrink, which is called a hook phase. The hook phase
ends with hydrogen ignition on the surface of the He core, and is
followed by a Hertzsprung gap (HG) phase in which the He core
continues to shrink while the hydrogen envelope begins expanding. At
some point, helium is ignited in the He core, and a core helium
burning (CHeB) phase starts. In general, a massive star never becomes
a red giant branch (RGB) star before entering into a CHeB phase. In
the CHeB phase, the stellar envelope transiently shrinks and expands
again, if the star is relatively light. This behavior is called a blue
loop. The stellar envelope monotonically expands if the star is
relatively heavy. The CHeB phase finishes when helium is completely
converted to carbon and oxygen (CO) at the center, and the CO core
emerges at the center. Subsequently, helium keeps burned on the
surface of the CO core. Thus, this phase is called a shell helium
burning (ShHeB) phase. The star becomes a RSG star in either of the
CHeB or ShHeB phase. The ShHeB phase continues until carbon is ignited
at the center. Shortly after the carbon ignition, the star experiences
a SN explosion, or gravitational collapse. Then, it finally leaves a
NS or BH.

There are three different points between Pop.~I/II stars and EMP stars
(including Pop.~III stars). (1) Some of EMP stars never become RSG
stars. (2) EMP stars experience smaller blue loops than Pop.~I/II
stars, and a part of EMP stars have no blue loops. (3) A part of EMP
stars skip HG phases. These points will be described in detail below.

\begin{figure*}
  \includegraphics[width=2\columnwidth,bb=0 0 320 116]{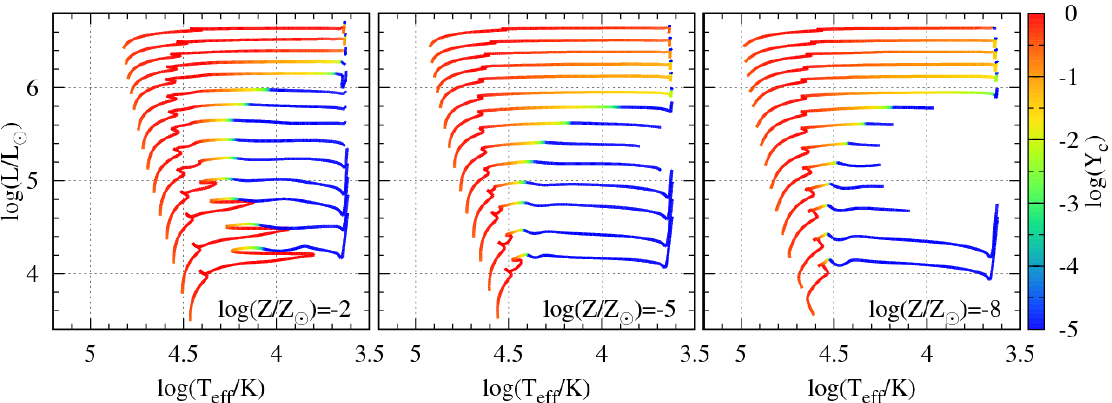}
  \caption{HR diagrams for stellar models with $\logz=-2, -5$, and
    $-8$. In each panel, curves indicate stellar evolutions with
    $M/M_\odot=8, 10, 13, 16, 20, 25, 32, 40, 50, 65, 80, 100, 125$,
    and $160$ from bottom to top. Colors are coded by the helium mass
    fractions in the stellar cores.
\label{fig:hrd}}
\end{figure*}

Figure~\ref{fig:hrd} shows Hertzsprung-Russell (HR) diagrams for
different metallicities. All the stars become RSG stars with $\teff
\lesssim 3.7$ for $\logz=-2$, while some of stars end with BSG stars
with $\teff \gtrsim 3.7$ for $\logz=-5$ and $-8$. The mass range of
stars ending with BSG stars becomes wider with metallicity decreasing:
$20 \lesssim M/M_\odot \lesssim 32$ for $\logz=-5$, and $13 \lesssim
M/M_\odot \lesssim 40$ for $\logz=-8$. This is because stars have
smaller opacity as they become more metal-poor.  Global features of
the HR diagram are not changed when the resolution of the evolution
calculation is increased.  The time evolution of surface temperature
between HG and RSG phases and the variation of luminosity in RSG phase
are slightly affected by the resolution in some cases.

\begin{figure*}
   \includegraphics[width=2\columnwidth,bb=0 0 348 116]{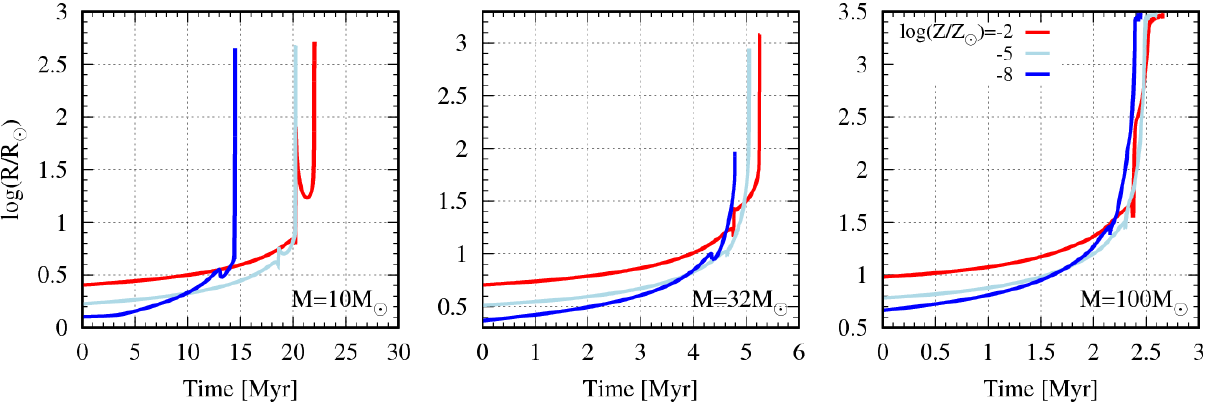}
  \caption{Radius evolutions of stars with $M/M_\odot=10$, $32$, and
    $100$ for $\logz=-2, -5$, and $-8$.\label{fig:radius}}
\end{figure*}

In Figure~\ref{fig:hrd}, we can see the absence of blue loops for EMP
stars. For $\logz = -2$, relatively light stars ($8 \lesssim M/M_\odot
\lesssim 25$) have blue loops. The $M=8M_\odot$ star has the most
prominent one among them. Its effective temperature returns up to
$\teff \sim 4.3$ after its effective temperature decreases down to
$\teff \sim 3.8$ once. For $\logz = -5$, stars with $8 \lesssim
M/M_\odot \lesssim 16$ still have blue loops, and however their blue
loops are much less prominent than those of stars for $\logz =
-2$. Even for the $M=8M_\odot$ star, the beginning temperature of the
blue loop is different from the highest temperature in the blue loop
only by $\Delta \teff \lesssim 0.1$. Finally, blue loops disappear
from all the stars for $\logz = -8$.

EMP stars may skip their HG phases. This can be seen in
Figure~\ref{fig:radius} which shows the radius evolutions of stars
with different masses and metallicities. For a star with $M=10M_\odot$
and $\logz = -2$, its radius monotonically increases from $\logr \sim
0.4$ to $\logr \sim 0.8$ until $t \sim 20$~Myr in its MS phase,
slightly decreases by $\Delta \logr \sim 0.1$ in the hook phase, and
increases by $\Delta \logr \sim 1.1$ on a short timescale in the HG
phase. We can see that the radius also increases by $\Delta \logr \sim
0.2$ at $t \sim 18$~Myr in the HG phase for $\logz = -5$. However, the
increase of the radius for $\logz = -5$ is much smaller than for
$\logz = -2$. Such radius increase is absent for $\logz = -8$; the HG
phase disappears for $\logz = -8$. Although the stellar radius
increases for $\logz = -8$ at $t \sim 14$~Myr, the star have entered
into the ShHeB phase at this time. From the above, we see the presence
and absence of the HG phases for the case of $M=10M_\odot$. This can
be true for the case of $M=32$ and $100M_\odot$ (see the middle and
right panels of Figure~\ref{fig:radius}). More metal-poor stars have
higher temperature at their centers at their MS phases due to
inefficiency of the CNO cycle. Then, their central temperature more
easily exceeds the temperature of helium ignition during their hook
phases. The dependence of the beginning time of CHeB phases is
systematically discussed in \cite{2019A&A...625A.132S}.

We should remark difference between Hurley's model and our model for
$\logz = -2$. In Hurley's model, which supports for $-2 \lesssim \logz
\lesssim 0$, the stars always become RSG stars when their CHeB phases
end. On the other hand, in our model, relatively light stars still
remain BSG stars when their CHeB phases end. This can be seen in
Figure~\ref{fig:hrd}. For $\logz = -2$, stars with $M/M_\odot \lesssim
50$ still remain BSG stars even when their central helium mass
fractions are decreased down to less than $10^{-5}$; they finish their
CHeB phases.

In summary, our model has three different points from Hurley's model
due to lower metallicity. Some of EMP stars end with BSG stars, and
skip HG and blue loop phases. There is one different point between
Hurley's and our model even for Pop.~I/II stars ($\logz = -2$). Stars
in Hurley's model necessarily become RSG stars when they finish their
CHeB phases, while stars in our model may remain BSG stars when they
finish their CHeB phases.

Finally, we compare our model of $\logz = -8$ with a zero-metal model
of \cite{2001A&A...371..152M}. Here, we identify our model of $\logz =
-8$ with a zero-metal model. In our model, stars with $13 \le
M/M_\odot \le 40$ end with BSG stars, and stars with $M/M_\odot \le
10$ or $M/M_\odot \ge 50$ end with RSG stars. Thus, the lower and
upper mass limits of stars ending with BSG stars are $10 < M/M_\odot
\le 13$ and $40 < M/M_\odot \le 50$, respectively. On the other hand,
in Marigo's model, the corresponding mass limits are $9.5 < M/M_\odot
\le 10$ and $50 < M/M_\odot \le 70$, respectively. The mass range of
stars ending with BSG stars in our model is in a good agreement with
that in Marigo's model, although our mass range is slightly smaller
than Marigo's.

\section{Implementation}
\label{sec:implementation}

In this section, we show the way to devise fitting formulae for
evolution tracks of massive EMP stars, using our stellar evolution
models as reference. The fitting formulae consist of a luminosity,
radius, and He core mass as functions of time ($t$), mass ($M$), and
metallicity ($Z$). We never construct three variable functions for
these quantities. Instead, we develop bivariate functions with
different metallicity, $\logz = -2, -4, -5, -6$ and $-8$. The
bivariate functions have totally different forms among stellar
evolution phases, since stars evolve differently among these
phases. We divide a stellar evolution into five phases: MS, HG, CHeB,
ShHeB, and remnant phases. Note that the MS phase includes a hook
phase, and massive stars skip RGB phases. We construct a bivariate
function for a stellar quantity for a given phase as follows. We make
fitting formulae for the stellar quantities at the beginning and
ending times of the phase as a function of $M$. We obtain the stellar
quantity at a given time of the phase by a simple polynomial
interpolation that bridges the stellar quantities at the beginning and
ending times of the phase.

\begin{table}
  \centering
  \caption{Mass limits constraining stellar evolutions.}
  \label{tab:massLimit}
  \begin{tabular}{cccccc}
    $\logz$ & $-2$ & $-4$ & $-5$ & $-6$ & $-8$ \\
    \hline
    $\mhgu$ &  --   &  --   & $25$ & $10$ &  $8$ \\
    $\mblu$ &  $32$ &  $32$ & $20$ & $10$ &  $8$ \\
    $\mebl$ &  $25$ &  $25$ & $20$ & $16$ & $13$ \\
    $\mebu$ &  $25$ &  $25$ & $40$ & $50$ & $50$ \\
    $\mcbu$ & $100$ &  $50$ & $50$ & $50$ & $50$ \\
    \hline
  \end{tabular}
\end{table}

As described in section~\ref{sec:stellarEvolutionModels}, the critical
masses of stars experiencing HG phases and blue loops become smaller
with metallicity decreasing. Moreover, the mass range of stars ending
with BSG stars is extended with metallicity decreasing. We consider
such metallicity dependences by defining five mass limits: the upper
mass limit of stars entering into HG phases ($\mhgu$), the upper mass
limit of stars with blue loops ($\mblu$), the upper and lower mass
limits of stars ending with BSG stars ($M_{\rm EB,u}$ and $\mebl$,
respectively), and the upper mass limit of stars remaining BSG stars
in CHeB phases ($\mcbu$). Then, stars enter into HG phases if $M <
\mhgu$, and have blue loops if $M < \mblu$. We indicate $\mhgu$ as
``--'' for $\logz = -2$ and $-4$, since all the stars enter into HG
phases for these metallicities. Stars end with BSG stars if $\mebl \le
M < \mebu$. Stars remain BSG stars in their CHeB phases if $M <
\mcbu$. We summarize these mass limits in Table~\ref{tab:massLimit}.

We can categorize phases (MS, HG, CHeB, ShHeB, remnant (NS/BH), BSG,
and RSG) into two types. The first type includes MS, HG, CHeB, ShHeB,
and remnant phases, which indicate states of stellar cores. The second
type contains BSG and RSG phases indicating stellar surfaces. The
former and latter types of phases are not exclusive. For example, a
star can be in CHeB and BSG phases. We summarize these phases in
Table~\ref{tab:DefinitionOfPhase}.

\begin{table*}
  \centering
  \caption{Definitions of phases.}
  \label{tab:DefinitionOfPhase}
  \begin{tabular}{ll}
    Phase & Definition \\
    \hline
    MS      & Phase from core hydrogen ignition to shell hydrogen ignition \\
    HG      & Phase from shell hydrogen ignition to core He ignition \\
    CHeB    & Phase from core He ignition to the end of core He burning \\
    ShHeB   & Phase from the end of core He burning to the carbon ignition \\
    Remnant & Phase in which a star do not evolve any more as a single star \\
    \hline
    BSG     & Phase with $\teff \ge 3.65$  \\
    RSG     & Phase with $\teff < 3.65$ \\
    \hline
  \end{tabular}
\end{table*}

We need to define these phases numerically in order to extract these
phases from our stellar evolution model, i.e. 1D simulation
data. These definitions are so technical, since we treat wide ranges
of stellar masses and metallicities. The definitions of the first type
of phases are as follows.
\begin{itemize}
\item MS phase. It starts from the ZAMS time defined as the time when
  a duration passes after the beginning of 1D simulation. The duration
  is 2 times of the maximum Kelvin-Helmholtz (KH) time in 1D
  simulations. The maximum KH time is achieved around the beginning
  time of the simulation. We can avoid a contraction phase from the
  initial hydrostatic equilibrium to the core hydrogen ignition. The
  MS phase ends at the time when the radius achieves the first local
  minimum.
\item HG phase. It starts from the ending time of the MS phase, and
  ends at the He ignition time.
\item CHeB phase. It starts at the He ignition time. If a star does
  not have an HG phase ($M \ge M_{\rm HG,u}$), the He ignition time is
  defined as the ending time of the MS phase. If a star have an HG
  phase ($M < M_{\rm HG,u}$), the time is defined as the time when the
  star stop the rapid increase of its radius (see
  Figure~\ref{fig:radius}). Although the definition is not directly
  related to He ignition, He is ignited in the core around at that
  time. The CHeB phase ends at the time when the central He mass
  fraction is decreased down to $10^{-3}$.
\item ShHeB phase. It starts from the ending time of the CHeB phase,
  and ends at the carbon ignition time, i.e. the ending time of 1D
  simulation.
\end{itemize}

In our definitions, stars with $160M_\odot$ enter into CHeB phases at
$\teff \sim 4$ for $\logz = -5$ and $-8$, while it does at $\teff \sim
3.7$ for $\logz = -2$ (see also Figure~\ref{fig:comparison}). It seems
inconsistent with Figure~\ref{fig:hrd}, since their He mass fractions
evolve similarly. However, it is due to a logarithmic color code. Our
definitions are actually consistent with Figure~\ref{fig:hrd}. The
former stars burn He, decreasing the He mass fractions from $\sim 1$
to $\sim 0.4$ until their $\teff$ go down to $\sim 3.65$. On the other
hand, the latter star does not burn He, keeping the He mass fractions
$\gtrsim 0.9$ when it is a BSG phase.

The definitions of the second type of phases are as follows.
\begin{itemize}
\item BSG phase. It starts from the ZAMS time. Its ending time is when
  $\teff$ becomes smaller than $3.65$.
\item RSG phase. It starts at the ending time of the BSG phase, and
  ends at the carbon ignition time, i.e. the ending time of 1D
  simulation.
\end{itemize}

Although stars with $\teff \sim 3.7$ are usually classified as yellow
supergiants \citep[e.g.][]{2009ApJ...703..441D}, we include them as
BSG phases in the same way as the \bse~code. In the \bse~code, stars
with radiative and convective envelopes are called stars in BSG and
RSG phases, respectively.

For supplement, we also show the dependence of stellar evolutionary
paths on stellar masses in Table~\ref{tab:EvolutionaryPath}. We omit
MS and HG phases, since they are always BSG stars in our fitting
formulae. As seen in Table~\ref{tab:massLimit}, $M_{\rm EB,l}=M_{\rm
  EB,u}$ for $\logz = -2$ and $-4$. This means all the stars become
RSG stars for $\logz = -2$ and $-4$, and no star applies to the second
line in Table~\ref{tab:EvolutionaryPath}. For $\logz = -6$ and $-8$,
$M_{\rm EB,u}=M_{\rm CB,u}$. Thus, no star applies to the third line
in Table~\ref{tab:EvolutionaryPath}.

\begin{table}
  \centering
  \caption{Stellar evolutionary paths.}
  \label{tab:EvolutionaryPath}
  \begin{tabular}{lll}
                                        & CHeB                   & ShHeB \\
    \hline
    $M < M_{\rm EB,l}$                  & BSG                    & BSG $\rightarrow$ RSG \\
    $M_{\rm EB,l} \le M < M_{\rm EB,u}$ & BSG                    & BSG \\
    $M_{\rm EB,u} \le M < M_{\rm CB,u}$ & BSG                    & BSG $\rightarrow$ RSG \\
    $M_{\rm CB,u} \le M$                & BSG $\rightarrow$ RSG  & RSG \\
    \hline
  \end{tabular}
\end{table}
  
Some stars can evolve to naked He stars when their hydrogen envelopes
are stripped by stellar winds or Roche-lobe overflow. Then, we instead
use fitting formulae of naked He stars developed by
\cite{2000MNRAS.315..543H}: $Z=0.0002$ naked He stars for Hurley's
fitting formulae with $\logz=-2$, and $Z=0.0001$ naked He stars for
Hurley's fitting formulae with $\logz \le -2$. Note that $Z=0.0001$ is
the lowest metallicity among Hurley's fitting formulae. Thus, in our
fitting formulae, naked He stars evolve differently among different
$\logz$ only if we include stellar wind mass loss.

Stars can have high spin velocities initially, and can be spun up by
tidal interactions with their companion stars. Then, they experience
chemically homogeneous evolution
\citep{2009A&A...497..243D}. Chemically homogeneous evolution may play
important roles in the formation of merging BH-BHs
\citep{2016A&A...588A..50M,2016MNRAS.458.2634M}. \cite{2012A&A...542A.113Y}
have studied chemically homogeneous evolution of Pop.~III stars.
However, we do not implement fitting formulae considering the
chemically homogeneous evolution. This is beyond the scope of this
paper.

In the following sections, we describe bivariate functions for
luminosity, radius, and He core mass in MS
(section~\ref{sec:MSPhase}), HG (section~\ref{sec:HGPhase}), CHeB
(section~\ref{sec:CHeBPhase}), ShHeB (section~\ref{sec:ShHeBPhase}),
and remnant phases (section~\ref{sec:RemnantPhase}). Since we
demonstrate our fitting formulae coupled with stellar wind mass loss
in section~\ref{sec:demonstration}, we describe our treatment of
stellar wind mass loss in section~\ref{sec:stellarwindmodel}. Due to
stellar winds, post-MS stars sometimes transition to naked He
stars. We describe how to change post-MS stars to naked He stars in
section~\ref{sec:transitionToNakedHeStars}. We describe fitting
parameters in Appendix~\ref{sec:valuesForFittingFormula}.

\subsection{MS phase}
\label{sec:MSPhase}

In this phase, stars evolve their luminosities and radii, while they
remain their He core mass to be zero. Thus, we construct two bivariate
functions for stellar luminosities and radii. We use the hat symbol,
such that $\hat{X}=\log X$. The bivariate functions can be expressed
as
\begin{align}
  \hat{L}_{\rm MS} &= \hat{L}_{\rm ZAMS} + \alpha_{\rm L} \tau_{\rm
    MS} + \beta_{\rm L} \tau_{\rm MS}^{20} \nonumber \\
  &+ \left[ \hat{L}_{\rm EMS} - \hat{L}_{\rm ZAMS} - \alpha_{\rm L} -
    \beta_{\rm L} \right] \tau_{\rm MS}^2 - \Delta L \left( \tau_{\rm
    MS}^2 - \tilde{\tau}_{\rm MS}^2 \right) \label{eq:LuminosityMS} \\
  \hat{R}_{\rm MS} &= \hat{R}_{\rm ZAMS} + \alpha_{\rm R} \tau_{\rm
    MS} + \beta_{\rm R} \tau_{\rm MS}^{10} + \gamma_{\rm R} \tau_{\rm
    MS}^{40} \nonumber \\
  &+ \left[ \hat{R}_{\rm EMS} - \hat{R}_{\rm ZAMS} - \alpha_{\rm R} -
    \beta_{\rm R} - \gamma_{\rm R} \right] \tau_{\rm MS}^3 - \Delta R
  \left(\tau_{\rm MS}^3 - \tilde{\tau}_{\rm MS}^3
  \right). \label{eq:RadiusMS}
\end{align}
All the variables other than $\tau_{\rm MS}$ and $\tilde{\tau}_{\rm
  MS}$ in the right-hand sides of the above equations are functions of
$M$, and $\tau_{\rm MS}$ and $\tilde{\tau}_{\rm MS}$ are functions of
$t$ and $M$, and therefore $L_{\rm MS}$ and $R_{\rm MS}$ are functions
of $t$ and $M$. Hereafter, we show variables in the right-hand sides
of the above equations step by step.

We indicate a scaled time in the MS phase by $\tau_{\rm MS}$. The
definition of $\tau_{\rm MS}$ is given by
\begin{align}
  \tau_{\rm MS} = \frac{t}{t_{\rm EMS}}, \label{eq:taums}
\end{align}
where $t_{\rm EMS}$ is the ending time of the MS phase. We model
$t_{\rm EMS}$ as
\begin{align}
  t_{\rm EMS} = \left\{
  \begin{array}{ll}
    0.99 t_{\rm HeI} & (M < \mhgu) \\
    t_{\rm HeI} & (M \ge \mhgu)
  \end{array}
  \right., \label{eq:EMSTime}
\end{align}
where $t_{\rm HeI}$ is the He ignition time, or the beginning time of
a CHeB phase written in
section~\ref{sec:CHeBPhase}. Eq.~(\ref{eq:EMSTime}) means that stars
with $M \ge \mhgu$ skip HG phases.

We make fitting formulas for luminosity and radius at the ZAMS time
($L_{\rm ZAMS}$ and $R_{\rm ZAMS}$, respectively), and those at the
ending time of the MS phase ($L_{\rm EMS}$ and $R_{\rm EMS}$,
respectively), such that
\begin{align}
  &\hat{L}_{\rm ZAMS} = \sum_{i=0}^3 { \cal L}_{{\rm ZAMS},i}
  \hat{M}^i, \;
  \hat{R}_{\rm ZAMS} = \sum_{i=0}^3 {\cal R}_{{\rm ZAMS},i} \hat{M}^i,
  \\
  &\hat{L}_{\rm EMS} = \sum_{i=0}^3 {\cal L}_{{\rm EMS},i} \hat{M}^i,
  \;
  \hat{R}_{\rm EMS} = \sum_{i=0}^3 {\cal R}_{{\rm EMS},i} \hat{M}^i,
\end{align}
where the coefficients ${\cal L}$ and ${\cal R}$ in the right-hand
sides of the above equations are constants for a given metallicity,
shown in section~\ref{sec:valuesForFittingFormula}. We relate
coefficients ${\cal L}$ and ${\cal R}$ to luminosities and radii,
respectively. We also make fitting formulas for Greek coefficients in
Eqs.~(\ref{eq:LuminosityMS}) and (\ref{eq:RadiusMS}). These can be
written as
\begin{align}
  &\alpha_{\rm L} = \sum_{i=0}^{3} {\cal L}_{\alpha, i} \hat{M}^{i-1},
  \; \beta_{\rm L} = \sum_{i=0}^{3} {\cal L}_{\beta, i} \hat{M}^{i-1}
  \\
  &\alpha_{\rm R} = \sum_{i=0}^{3} {\cal R}_{\alpha, i} \hat{M}^{i-1},
  \; \beta_{\rm R} = \sum_{i=0}^{3} {\cal R}_{\beta, i} \hat{M}^{i-1},
  \; \gamma_{\rm R} = \sum_{i=0}^{3} {\cal R}_{\gamma, i}
  \hat{M}^{i-1}.
\end{align}

Eqs.~(\ref{eq:LuminosityMS}) and (\ref{eq:RadiusMS}) contain terms
with $\Delta L$ and $\Delta R$, respectively. These terms consider
drastic brightening and shrinkage in the hook phase. The variable
$\tilde{\tau}_{\rm MS}$ can be written as
\begin{align}
  \tilde{\tau}_{\rm MS} &= \max \left\{ 0.0, \min \left[ 1.0,
    \frac{\tau_{\rm MS} - (1.0 - \epsilon)}{\epsilon } \right]
  \right\}
\end{align}
for $\epsilon = 0.01$. We can see that $\tilde{\tau}_{\rm MS}$
suddenly increases from 0 to 1 during $1-\epsilon<\tau_{\rm MS}<1$, or
in the hook phase. The correction terms ($\Delta L$ and $\Delta R$)
can be expressed as
\begin{align}
  \Delta L = \sum_{i=0}^{3} {\cal L}_{\Delta,i} \hat{M}^{i}, \;\;
  \Delta R = \sum_{i=0}^{3} {\cal R}_{\Delta,i} \hat{M}^{i}.
\end{align}

The fitting formulae Eq.~(\ref{eq:LuminosityMS}) and
(\ref{eq:RadiusMS}) have the most complex forms and the largest
numbers of parameters among our fitting formulae. Because of this,
they match well the reference stellar models, dealing with drastic
changes near the end of the MS phase.

\subsection{HG phase}
\label{sec:HGPhase}

Stars enter into these phases when $M < \mhgu$. Their luminosity and
radius can be written as
\begin{align}
  \hat{L}_{\rm HG} &= \hat{L}_{\rm EMS} + \tau_{\rm HG} \left(
  \hat{L}_{\rm HeI} - \hat{L}_{\rm EMS}
  \right) \label{eq:LuminosityHG} \\
  \hat{R}_{\rm HG} &= \hat{R}_{\rm EMS} + \tau_{\rm HG} \left(
  \hat{R}_{\rm HeI} - \hat{R}_{\rm EMS} \right), \label{eq:RadiusHG}
\end{align}
where $L_{\rm HeI}$ and $R_{\rm HeI}$ are the luminosity and radius at
the He ignition time shown in section~\ref{sec:CHeBPhase}. We define
a scaled time in the HG phase, $\tau_{\rm HG}$, as
\begin{align}
  \tau_{\rm HG} = \frac{t-t_{\rm EMS}}{t_{\rm HeI}-t_{\rm
      EMS}}. \label{eq:tauhg}
\end{align}

Stars first have non-zero He core mass in their HG phases. The
evolutions of the He core mass can be expressed as
\begin{align}
  M_{\rm c,HG} = \left[ (1 - \tau_{\rm HG}) M_{\rm c,HG,i} + \tau_{\rm
      HG} \right] M_{\rm c,HeI}, \label{eq:HecoreHG}
\end{align}
where $M_{\rm c,HG,i}$ and $M_{\rm c,HeI}$ are the He core mass at the
beginning time of the HG phase, and at the He ignition time,
respectively. We set $M_{\rm c,HG,i}$ in the same as
\cite{2000MNRAS.315..543H}:
\begin{align}
  M_{\rm c,HG,i} = \frac{1.586 + M^{5.25}}{2.434 + 1.02M^{5.25}}. \label{eq:HecoreHGi}
\end{align}
Although there is no reason for agreement between
Eq.~(\ref{eq:HecoreHGi}) and our stellar models, their deviations are
a few \% at most. Thus, we adopt it. The fitting formula of $M_{\rm
  c,HeI}$ is described in section~\ref{sec:CHeBPhase}.

The fitting formulae Eq.~(\ref{eq:LuminosityHG}), (\ref{eq:RadiusHG}),
and (\ref{eq:HecoreHG}) have quite simple time
interpolation. Actually, they are the same as in those of
\cite{2000MNRAS.315..543H}. They match well the reference stellar
models. This is because values at the beginning and ending times of
this phase are in a good agreement with each other, and because the
timescale of this phase is quite short.

\subsection{CHeB phase}
\label{sec:CHeBPhase}

We make bivariate functions for He core mass, luminosity, and radius
in this phase. The function of the He core mass ($M_{\rm c,CHeB}$) is
written as
\begin{align}
  M_{\rm c,CHeB} = M_{\rm c,HeI} + (M_{\rm c,ECHeB} - M_{\rm c,HeI})
  \tau_{\rm CHeB}, \label{eq:HecoreCHeB}
\end{align}
where $M_{\rm c,HeI}$ and $M_{\rm c,ECHeB}$ are the He core mass at
the He ignition time and the ending time of the CHeB phase,
respectively, and $\tau_{\rm CHeB}$ is a scaled time in this phase. We
define $\tau_{\rm CHeB}$, such that
\begin{align}
  \tau_{\rm CHeB} &= \frac{t-t_{\rm HeI}}{t_{\rm CHeB}},
\end{align}
where $t_{\rm CHeB}$ is the time interval of the CHeB phase. We make
the fitting formulas of $t_{\rm HeI}$ and $t_{\rm CHeB}$, such that
\begin{align}
  t_{\rm HeI} &= \sum_{i=0}^3 {\cal T}_{{\rm HeI},i}
  M^{-i}. \label{eq:HeITime} \\
  t_{\rm CHeB} &= \sum_{i=0}^3 {\cal T}_{{\rm CHeB},i} M^{-i},
\end{align}
respectively. Note that we relate coefficients ${\cal T}$ to the
beginning and ending times of a phase, and the time interval of a
phase.  The He core masses at the He ignition time and the ending time
of the CHeB phase are expressed as
\begin{align}
  \hat{M}_{\rm c,HeI} &= \sum_{i=0}^3 {\cal H}_{{\rm HeI},i} \hat{M}^i
  \\
  \hat{M}_{\rm c,ECHeB} &= \sum_{i=0}^3 {\cal H}_{{\rm ECHeB},i}
  \hat{M}^i.
\end{align}
Note that we relate coefficients ${\cal H}$ to He core masses.

The function for luminosities in this phase can be obtained by the
following equation:
\begin{align}
  \hat{L}_{\rm CHeB} &= \hat{L}_{\rm HeI} + \lambda \left(
  \hat{L}_{\rm ECHeB} - \hat{L}_{\rm HeI}
  \right), \label{eq:LuminosityCHeB} \\
  \lambda &= \tau_{\rm CHeB}^\xi, \label{eq:lambda} \\
  \xi &= \min \left[ 2.5, \max \left( 0.4, R_{\min}/R_{\rm HeI}
    \right) \right]. \label{eq:xi}
\end{align}
The luminosities at the He ignition time ($L_{\rm HeI}$), and the
ending time of the CHeB phase ($L_{\rm ECHeB}$) are given by
\begin{align}
  \hat{L}_{\rm HeI} &= \sum_{i=0}^3 {\cal L}_{{\rm HeI},i} \hat{M}^i
  \\
  \hat{L}_{\rm ECHeB} &= \sum_{i=0}^3 {\cal L}_{{\rm ECHeB},i}
  \hat{M}^i.
\end{align}
The index $\xi$ contains the minimum radius in the CHeB phase
($R_{\min}$), and the radius at the He ignition time, written below in
detail.

The function for radii in this phase depends on whether stars are BSG
or RSG stars, such that
\begin{align}
  \hat{R}_{\rm CHeB} = \left\{
  \begin{array}{ll}
    \hat{R}_{\min} + |\rho|^3 & (\tau_{\rm CHeB} \le \tau_{\rm
      CHeB,EBSG}) \\
    \hat{R}_{\rm CHeB,RSG} & (\tau_{\rm CHeB} > \tau_{\rm
      CHeB,EBSG}) \\
  \end{array}
  \right.. \label{eq:RadiusCHeB}
\end{align}
Note that $R_{\rm CHeB,RSG}$ is radii of RSG stars in this phase as
functions of $t$ and $M$, described later in detail. We indicate
$\tau_{\rm CHeB,EBSG}$ as the scaled time when a star finishes its BSG
or CHeB phase. Thus, $\tau_{\rm CHeB,EBSG}$ can be expressed as
\begin{align}
  \tau_{\rm CHeB,EBSG} &= \frac{\min(t_{\rm EBSG},t_{\rm CHeB}+t_{\rm
      HeI})-t_{\rm HeI}}{t_{\rm CHeB}},
\end{align}
where $t_{\rm EBSG}$ is the time when a star finishes its BSG
phase. We can express $t_{\rm EBSG}$ as
\begin{align}
  t_{\rm EBSG} = \left\{
  \begin{array}{ll}
    \displaystyle \sum_{i=0}^3 {\cal T}_{{\rm EBSG},i} M^{-i} & (M <
    \mebl, \; \mbox{or}\; \mebu \le M) \\
    t_{\rm Fin} & (\mebl \le M < \mebu) \\
  \end{array}
  \right.,
\end{align}
where $t_{\rm Fin}$ is the time when a star finishes its life,
described in detail in section~\ref{sec:ShHeBPhase}. Since star with
$\mebl \le M < \mebu$ ends their lives with BSG stars, $t_{\rm EBSG} =
t_{\rm Fin}$.

We first show a radius of a RSG star in this phase. The radius
explicitly depends not on $M$ but on $L_{\rm CHeB}$, such that
\begin{align}
  \hat{R}_{\rm CHeB,RSG} = \sum_{i=0}^1 {\cal R}_{{\rm RSG},i}
  \hat{L}_{\rm CHeB}^i. \label{eq:RadiusCHeBRSG}
\end{align}
The coefficients ${\cal R}_{{\rm RSG,}i}$ are functions of $M$, given
by
\begin{align}
  {\cal R}_{{\rm RSG},i} = \sum_{j=0}^1 {\cal R}_{{\rm RSG},ij}
  \hat{M}^j.
\end{align}

We next explain radii of BSG stars. The minimum radius in this phase
($R_{\min}$) depends on whether a star has its blue loop or not. If a
star does not have its blue loop, its minimum radius in this phase is
equal to the radius at the He ignition time ($R_{\rm HeI}$). Thus,
we can write $R_{\rm HeI}$ and $R_{\min}$ as
\begin{align}
  R_{\rm HeI} &= \sum_{i=0}^3 {\cal R}_{{\rm HeI},i} \hat{M}^i, \\
  R_{\min} &= \left\{
  \begin{array}{ll}
    \displaystyle \sum_{i=0}^3 {\cal R}_{{\rm \min},i} \hat{M}^i & (M
    < \mblu) \\ 
    R_{\rm HeI} & (M \ge \mblu) \\
  \end{array}
  \right..
\end{align}
The increment of the radii at the BSG phase ($\rho$) is given by
\begin{align}
  \rho &= \left( \hat{R}_{\rm CHeB,EBSG} - \hat{R}_{\min}
  \right)^{1/3} \left( \frac{\tau_{\rm CHeB}}{\tau_{\rm CHeB,EBSG}}
  \right) \nonumber \\
  &- \left( \hat{R}_{\rm HeI}- \hat{R}_{\min} \right)^{1/3} \left( 1 -
  \frac{\tau_{\rm CHeB}}{\tau_{\rm CHeB,EBSG}} \right), \label{eq:rho}
\end{align}
where $R_{\rm CHeB,EBSG}$ is the radius at $\tau = \tau_{\rm
  EBSG}$. We express $R_{\rm CHeB,EBSG}$ as
\begin{align}
  \hat{R}_{\rm CHeB,EBSG} = \left\{
  \begin{array}{ll}
    \displaystyle \sum_{i=0}^3 {\cal R}_{{\rm CHeB,EBSG},i} 
      \hat{M}^i & (M < \mcbu) \\
   \hat{R}_{\rm CHeB,RSG,EBSG}& (M \ge \mcbu) \\
  \end{array}
  \right.. 
\end{align}
We properly make a fitting formula of $R_{\rm CHeB,EBSG}$ for $M <
\mcbu$, and otherwise use Eq.~(\ref{eq:RadiusCHeB}) for $\tau_{\rm
  CHeB} = \tau_{\rm EBSG}$.

The fitting formulae in this phase (Eq.~(\ref{eq:HecoreCHeB}),
(\ref{eq:LuminosityCHeB}), and (\ref{eq:RadiusCHeB})) have the same
interpolations as those of \cite{2000MNRAS.315..543H}, except for that
of a radius in a RSG phase. They match well the reference stellar
models. This is because important values are consistent with each
other. The important values are those at the beginning and ending
times of this phase for He core mass and luminosity. For a radius of a
BSG phase, the minimum value is also important as well as values at
the beginning and ending times. The fitting formula of a radius of a
RSG star fits to the reference stellar models, since the radius is
well correlated to the luminosity.

\subsection{ShHeB phase}
\label{sec:ShHeBPhase}

In this phase, we stop the evolution of the He core mass. Thus, the He
core mass remains constant as:
\begin{align}
  M_{\rm c,ShHeB} = M_{\rm c,ECHeB}. \label{eq:HecoreShHeB}
\end{align}
We simplify the evolution of the CO core mass, such that
\begin{align}
  \hat{M}_{\rm c,CO} = f_{\rm CO} \sum_{i=0}^3 {\cal C}_{i} \left[
    \log(M) \right]^i, \label{eq:COcoreShHeB}
\end{align}
where
\begin{align}
  f_{\rm CO} = \left\{
  \begin{array}{ll}
    0.99 & (t < t_{\rm Fin}) \\
    1.0  & (t = t_{\rm Fin}) \\
  \end{array}
  \right..
\end{align}
Note that coefficients ${\cal C}$ are related to CO core masses.  The
ending time of the stellar evolution is expressed as
\begin{align}
  t_{\rm Fin} = \sum_{i=0}^3 {\cal T}_{{\rm Fin},i} M^{-i}.
\end{align}
The CO core mass is used for calculating remnant mass described in
section~\ref{sec:RemnantPhase}.

The function of luminosities in this phase can be divided according to
a BSG or RSG star, such that
\begin{align}
  \hat{L}_{\rm ShHeB} = \left\{
  \begin{array}{ll}
    \hat{L}_{\rm ECHeB} - \tau_{\rm BSG}^3 \left( \hat{L}_{\rm EBSG} -
    \hat{L}_{\rm ECHeB} \right) & (t \le t_{\rm EBSG}) \\
    \hat{L}_{\rm EBSG} - \tau_{\rm RSG} \left( \hat{L}_{\rm Fin} -
    \hat{L}_{\rm EBSG} \right) & (t > t_{\rm EBSG}) \\
  \end{array}
  \right., \label{eq:LuminosityShHeB}
\end{align}
where $\tau_{\rm BSG}$ and $\tau_{\rm RSG}$ are scaled times in BSG
and RSG phases, and expressed as
\begin{align}
  \tau_{\rm BSG} &= \frac{t-(t_{\rm HeI}+t_{\rm CHeB})}{t_{\rm
      EBSG}-(t_{\rm HeI}+t_{\rm CHeB})} \\ 
  \tau_{\rm RSG} &= \frac{t-t_{\rm EBSG}}{t_{\rm Fin}-t_{\rm EBSG}},
\end{align}
respectively. The functional form of $L_{\rm EBSG}$ is bifurcated by
whether the star ends its life with a BSG or RSG star, and is given by
\begin{align}
  \hat{L}_{\rm EBSG} = \left\{
  \begin{array}{ll}
    \displaystyle \sum_{i=0}^3 {\cal L}_{{\rm EBSG},i} \hat{M}^i & (M
    < \mebl, \; \mbox{or} \; \mebu \le M) \\
    \hat{L}_{\rm Fin} & (\mebl \le M < \mebu) \\
  \end{array}
  \right.,
\end{align}
where $L_{\rm Fin}$ is the luminosity at the ending time of the
evolution. Since stars with $\mebl \le M < \mebu$ end with BSG stars,
$L_{\rm EBSG}=L_{\rm Fin}$. We make a fitting formula for the
luminosity at the ending time of the evolution ($L_{\rm Fin}$), such
that
\begin{align}
  \hat{L}_{\rm Fin} = \left\{
  \begin{array}{ll}
    \sum_{i=0}^3 {\cal L}_{{\rm Fin,l},i} \hat{M}^i & (M < \mebl) \\
    \sum_{i=0}^3 {\cal L}_{{\rm Fin,u},i} \hat{M}^i & (M \ge \mebl)
  \end{array}
  \right..
\end{align}

The function for a radius in this phase also depends on whether the
star is in a BSG or RSG phase. Thus, we can write the function as
\begin{align}
  \hat{R}_{\rm ShHeB} = \left\{
  \begin{array}{ll}
    \hat{R}_{\rm ECHeB} - \tau_{\rm BSG}^3 (\hat{R}_{\rm EBSG} -
    \hat{R}_{\rm ECHeB}) & (t \le t_{\rm EBSG}) \\
    \displaystyle \sum_{i=0}^1 {\cal R}_{{\rm RSG},i} \hat{L}_{\rm
      ShHeB}^i & (t > t_{\rm EBSG}) \\
  \end{array}
  \right.. \label{eq:RadiusShHeB}
\end{align}
Note that ${\cal R}_{{\rm RSG},i}$ in the second expression in the
right-hand side of the above equation is the same as in
Eq.~(\ref{eq:RadiusCHeBRSG}). We can obtain the radius at the ending
time of the CHeB phase ($R_{\rm ECHeB}$) by using
Eq.~(\ref{eq:RadiusCHeB}) for $\tau_{\rm CHeB}=1$. The radius at the
ending time of a BSG phase ($R_{\rm EBSG}$) is expressed as
\begin{align}
  \hat{R}_{\rm EBSG} = \left\{
  \begin{array}{ll}
    \hat{R}_{\rm ShHeB,EBSG} & (M < \mebl, \; \mbox{or} \; \mebu \le
    M) \\
    \hat{R}_{\rm Fin} & (\mebl \le M < \mebu) \\
  \end{array}
  \right.,
\end{align}
where $R_{\rm ShHeB,EBSG}$ is $R_{\rm ShHeB}$ at $t=t_{\rm EBSG}$ in
Eq.~(\ref{eq:RadiusShHeB}), and $R_{\rm Fin}$ is the radius at the
ending time of the evolution. The above equation is bifurcated by
whether the star ends with a RSG (top) or BSG star (bottom). The
radius at the ending time of the evolution can be written as
\begin{align}
  \hat{R}_{\rm Fin} = \sum_{i=0}^3 {\cal R}_{{\rm Fin},i} \hat{M}^i.
\end{align}

We fix a He core mass in this phase in the same way as
\cite{2000MNRAS.315..543H}. This approximation is sufficient, since
the He core mass changes by at most $1$~\% in this phase in the
reference stellar models. We almost fix a CO core mass in this
phase. The CO core mass grows by at most $5$~\% for $M \lesssim
10M_\odot$, and $1$~\% for $M \gtrsim 10M_\odot$ in this phase in the
reference stellar models. Although we slightly overestimate the CO
core mass for $M \lesssim 10M_\odot$, we overestimate the remnant mass
only by less than $1$~\% for the following reason. When we calculate
the remnant mass, we adopt the top equations in
Eqs.~(\ref{eq:Remnant}) and (\ref{eq:FeCore}), since $M_{\rm
  c,CO}/M_\odot \le 2.5$ for $M \lesssim 10M_\odot$. Even if $M_{\rm
  c,CO}$ contains an error of $5$~\%, $M_{\rm Fe-Ni}$ contains an
error of less than $1$~\% owing to the small contribution of the first
term of Eq.~(\ref{eq:FeCore}). The fitting formulae of luminosity and
radius in this phase (Eq.~(\ref{eq:LuminosityShHeB}) and
(\ref{eq:RadiusShHeB})) match well the reference stellar models.  For
luminosity and a radius in a BSG phase, this reason is that values at
the beginning and ending times of this phase are in a good agreement
with each other.  For luminosity and radius of a BSG phase, the index
of $\tau_{\rm BSG}$ is $3$, since their evolutions become more rapid
at a later time. For luminosity of a RSG phase, the index of
$\tau_{\rm RSG}$ is $1$, since its evolution keeps nearly
constant. For a radius in a RSG phase, the reason for good agreement
between the fitting formulae and reference stellar models is that the
radius is well correlated to the luminosity in the RSG phase.

\subsection{Remnant phase}
\label{sec:RemnantPhase}

Stars on our fitting formulae become NSs or BHs. We set their
luminosities and radii to be the same as in
\cite{2000MNRAS.315..543H}. For the remnant mass, we implement two
models. We take into account the effects of PPI and PI~SNe for one
model (w/ PI), and do not for the other model (w/o PI). For the w/o
PI, we adopt the same formula as in \cite{2002ApJ...572..407B}, which
is also adopted in \cite{2014MNRAS.442.2963K}. The remnant mass can be
expressed as
\begin{align}
  &M_{\rm rem} = \nonumber \\
&\left\{
  \begin{array}{ll}
    M_{\rm Fe-Ni} & (M_{\rm c,CO}/M_\odot \le 5) \\
    \displaystyle M_{\rm Fe-Ni} + \frac{M_{\rm c,CO}-5}{2.6}(M-M_{\rm
      Fe-Ni}) & (5 < M_{\rm c,CO}/M_\odot < 7.6) \\
    M & (7.6 \le M_{\rm c,CO}/M_\odot) \\
  \end{array}
  \right., \label{eq:Remnant}
\end{align}
where
\begin{align}
  M_{\rm Fe-Ni} = \left\{
  \begin{array}{ll}
    0.161767M_{\rm c,CO} + 1.067055 & (M_{\rm c,CO}/M_\odot \le 2.5) \\
    0.314154M_{\rm c,CO} + 0.686008 & (2.5 < M_{\rm c,CO}/M_\odot) \\
  \end{array}
  \right.. \label{eq:FeCore}
\end{align}
For the w/ PI, we reduce the remnant mass obtained from the above
equations, such that
\begin{align}
  M_{\rm rem,PI} = \left\{
  \begin{array}{ll}
    M_{\rm rem} & (M_{\rm c,He}/M_\odot \le 45, 135 < M_{\rm c,He}/M_\odot) \\
    45 & (45 < M_{\rm c,He}/M_\odot \le  65) \\
    0 & (65 < M_{\rm c,He}/M_\odot \le 135)
  \end{array}
  \right.. \label{eq:RemnantPI}
\end{align}
PPI and PI~SNe work in the ranges of $45<M_{\rm c,He}/M_\odot\le 65$
and $65<M_{\rm c,He}/M_\odot\le135$, respectively. These thresholds
are the same as adopted by \cite{2016A&A...594A..97B}. We do not
consider mass loss via neutrino emission during BH formation, however
we can do readily if required. Therefore, the remnant mass of PPI is
equal to the lower threshold of PPI. Although we simplify PPI, there
are several studies which investigate PPI in detail
\citep{2017ApJ...836..244W,2019ApJ...882...36M}.

We regard a remnant as an NS if $M_{\rm min,NS} \le M_{\rm rem} \le
M_{\rm max,NS}$ (or $M_{\rm min,NS} \le M_{\rm rem,PI} \le M_{\rm
  max,NS}$), and as BH otherwise. In this paper, we set $M_{\rm
  min,NS}=1.3M_\odot$ and $M_{\rm max,NS}=3M_\odot$ tentatively. If
$M_{\rm rem} < M_{\rm min,NS}$, the star is a white dwarf, and its
mass is obtained by a different formula. We do not show the formula,
since all the stars in our fitting formulae do not become a white
dwarf unless stellar wind mass loss is taken into account.

For demonstration, we tentatively implement spin magnitudes of BHs
\begin{align}
  \chi = \frac{p_1-p_2}{2} \tanh \left(p_3 - M_{\rm rem} \right) +
  \frac{p_1+p_2}{2},
\end{align}
which is the ``{\it collapse}'' model of
\cite{2018PhRvD..98h4036G}. Here, $p_i=0.86 \pm 0.06, 0.13 \pm 0.13,
29.5 \pm 8.5$, and we adopt the median values. In this model, low-mass
BHs ($M_{\rm rem}/M_\odot \lesssim 29.5$) have high spins ($\chi \sim
p_1 = 0.86$), and high-mass BHs ($M_{\rm rem}/M_\odot \gtrsim 29.5$)
have low spins ($\chi \sim p_2 = 0.13$). Although
\cite{2018PhRvD..98h4036G} have modeled spin directions, we do not
describe them. The spin directions are correlated to binary
interactions, whereas we focus on single star evolutions in this
paper.

\subsection{Stellar wind model}
\label{sec:stellarwindmodel}

We describe a stellar wind model used in
section~\ref{sec:demonstration}. The stellar wind model is described
in \cite{2000MNRAS.315..543H} with modifications of
\cite{2010ApJ...714.1217B} and \cite{2018arXiv181009721K}. The stellar
wind mass loss $\dot{M}$ is given by
\begin{align}
  \dot{M} = \left\{
  \begin{array}{ll}
    \max(\dot{M}_{\rm NJ},\dot{M}_{\rm OB}) & (\mbox{MS}) \\
    \max(\dot{M}_{\rm NJ},\dot{M}_{\rm OB},\dot{M}_{\rm
      R},\dot{M}_{\rm WR})+\dot{M}_{\rm LBV} & (\mbox{HG and CHeB}) \\
    \max(\dot{M}_{\rm NJ},\dot{M}_{\rm OB},\dot{M}_{\rm
      R},\dot{M}_{\rm WR},\dot{M}_{\rm VW})+\dot{M}_{\rm LBV} &
    (\mbox{ShHeB}) \\
    \max(\dot{M}_{\rm R},\dot{M}_{\rm WR}) & (\mbox{naked He stars})
  \end{array}
  \right.,
\end{align}
where $\dot{M}_{\rm NJ},\dot{M}_{\rm OB},\dot{M}_{\rm R},\dot{M}_{\rm
  WR},\dot{M}_{\rm VW}$, and $\dot{M}_{\rm LBV}$ are described
below. All of them are in the unit of
$M_\odot\mbox{yr}^{-1}$. $\dot{M}_{\rm NJ}$ is mass loss of luminous
stars, expressed as
\begin{align}
  \dot{M}_{\rm NJ} = \left\{
  \begin{array}{ll}
    0 & (L \le 4000L_\odot) \\
    & \\
    9.6 \times 10^{-15} (R/R_\odot)^{0.81} (L/L_\odot)^{1.24}
    \\ \times (M/M_\odot)^{0.16} (Z/Z_\odot)^{0.5} & (L > 4000L_\odot)
  \end{array}
  \right.
\end{align}
\citep{1990A&A...231..134N,1989A&A...219..205K}. $\dot{M}_{\rm OB}$ is
mass loss of hot massive hydrogen-rich stars, such that
\begin{align}
  \log(\dot{M}_{\rm OB}) = \left\{
  \begin{array}{ll}
    -6.388 \\ + 2.210 \log(L/10^5L_\odot) \\ - 1.339 \log(M/30M_\odot)
    \\ + 0.85 \log(Z/Z_\odot) \\ + 1.07 \log(T_{\rm eff}/20000K) &
    (1.25 \le T_{\rm eff}/10^4K \le 2.5) \\
    & \\
    -6.837 \\ + 2.194 \log(L/10^5L_\odot) \\ - 1.313 \log(M/30M_\odot)
    \\ + 0.85 \log(Z/Z_\odot) \\ + 0.933 \log(T_{\rm eff}/40000K) \\ -
    10.92 [\log(T_{\rm eff}/40000K)]^2 & (2.5 \le T_{\rm eff}/10^4K
    \le 5)
  \end{array}
  \right.
\end{align}
\citep{Vink01}. $\dot{M}_{\rm R}$ is mass loss of stars on the giant
branch and beyond, described as
\begin{align}
  \dot{M}_{\rm R} = 2 \times 10^{-13} (L/L_\odot) (R/R_\odot)
  (M/M_\odot)^{-1}
\end{align}
\citep{1978A&A....70..227K,1983ARA&A..21..271I}. $\dot{M}_{\rm WR}$ is
mass loss for Wolf-Rayet (WR) stars or naked He stars, written as
\begin{align}
  &\dot{M}_{\rm WR} = 10^{-13} (L/L_\odot)^{1.5} (Z/Z_\odot)^{0.86} (1
  - \mu) \label{eq:mwr} \\
  &\mu = [(M-M_{\rm c,He})/M] \min \{5.0,
  \max[1.2,(L/70000L_\odot)^{0.5}]\}. \label{eq:muwr}
\end{align}
The coefficient and dependence of $L$ and $Z$ in Eq.~(\ref{eq:mwr})
come from \cite{1998A&A...335.1003H} and
\cite{2005A&A...442..587V}. The term $(1-\mu)$ in Eq.~(\ref{eq:mwr})
corrects mass loss for post-MS stars with small hydrogen
envelopes. $\dot{M}_{\rm VW}$ is mass loss of stars on the asymptotic
giant branch, such that
\begin{align}
  &\log(\dot{M}_{\rm VW}) = \nonumber \\
  &\max \{ -11.4+0.0125[P_0-100\max(M/M_\odot-2.5,0)], \nonumber \\
  &\log(1.36 \times 10^{-9} (L/L_\odot)) \},
\end{align}
\citep{1993ApJ...413..641V}.  Here, $P_0$ is Mira pulsation period,
given by
\begin{align}
  \log(P_0/\mbox{day}) =
  \min(3.3,-2.07-0.9\log(M/M_\odot)+1.94\log(R/R_\odot).
\end{align}
Finally, $\dot{M}_{\rm LBV}$ is mass loss of luminous blue variable
(LBV) stars, expressed as
\begin{align}
  \dot{M}_{\rm LBV} = \left\{
  \begin{array}{ll}
    1.5 \times 10^{-4} & (L > 6 \times 10^5 L_\odot, \; \mbox{and} \;
    x_{\rm LBV}>1) \\
    0 & (\mbox{otherwise})
  \end{array}
  \right., \label{eq:dotMLBV}
\end{align}
\citep{1994PASP..106.1025H}, where $x_{\rm LBV} = 10^{-5} (R/R_\odot)
(L/L_\odot)^{0.5}$.

\cite{2010ApJ...714.1217B} have shown that this stellar wind model is
applicable to Pop.~I/II stars, not to EMP stars. Nevertheless, we
extrapolate this model to EMP stars for demonstration in this
paper. Constructing a stellar wind model for EMP stars is beyond the
scope of this paper. Since we include stellar winds by a
post-processing way, we can replace this model with another model
easily.

We change stellar parameters along with stellar wind mass loss in the
same way as in section 7.1 of \cite{2000MNRAS.315..543H}. We briefly
show this method here. When MS and HG stars lose their masses due to
stellar winds, their scaled times $\tau_{\rm MS}$ and $\tau_{\rm HG}$
(see Eq.~(\ref{eq:taums}) and (\ref{eq:tauhg}), respectively) are kept
constant, and their luminosities and radii are changed along with
their current masses. For a HG star, its He core mass is also changed
along with its current mass so as not to be decreased. For CHeB and
ShHeB stars, two types of masses are needed: the mass at the He
ignition time ($M_0$), and the current mass ($M_{\rm t}$). We replace
$M$ with $M_0$ for fitting formulae of their scaled times,
luminosities, and He and CO core masses. We use $M_{\rm t}$ for
fitting formulae of their radii instead of $M$. The response of a
naked He star is the same as that of an MS star if He is burned in the
core, and is the same as those of CHeB and ShHeB stars if He is not
burned in the core. Here, $M_0$ is the mass at the ending time of He
burning in the core.

\subsection{Transition to naked He stars}
\label{sec:transitionToNakedHeStars}

Some of Post-MS stars (HG, CHeB, and ShHeB stars) transition to naked
He stars due to stellar winds.  In order to model this transition, we
use the same method as in section 6.3 of
\cite{2000MNRAS.315..543H}. Here, we outline this method. When $\mu <
1$ (see Eq.~(\ref{eq:muwr}) for $\mu$), we perturb stellar
luminosities and radii as follows:
\begin{align}
  L_{\rm trans} &= L_{\rm nHe} \left( \frac{L_{\rm postMS}}{L_{\rm
      nHe}} \right)^s \\
  R_{\rm trans} &= R_{\rm nHe} \left( \frac{R_{\rm postMS}}{R_{\rm
      nHe}} \right)^r,
\end{align}
where $L_{\rm nHe}$ and $L_{\rm postMS}$ are luminosities of naked He
and post-MS stars, and $R_{\rm nHe}$ and $R_{\rm postMS}$ are radii of
naked He and post-MS stars. The indexes $s$ and $r$ are described in
eq.~(101) and (102) of \cite{2000MNRAS.315..543H}. We obtain $L_{\rm
  postMS}$ and $R_{\rm postMS}$ from formulae in
sections~\ref{sec:HGPhase}, \ref{sec:CHeBPhase}, and
\ref{sec:ShHeBPhase}, taking into account stellar wind mass loss
described in section~\ref{sec:stellarwindmodel}. We adopt Hurley's
fitting formulae of naked He stars with $Z=0.0002$ and $0.0001$ for
the cases of $\logz=-2$ and $\logz<-2$, respectively.

\section{Demonstration}
\label{sec:demonstration}

In this section, we demonstrate our fitting formulae. We use the
fitting formulae through the \sse\;code. In
section~\ref{sec:comparison}, we compare our fitting formulae with our
reference stellar models. Then, we do not take into account stellar
wind mass loss. In section~\ref{sec:combination}, we combine our
fitting formulae with a stellar wind model described in
section~\ref{sec:stellarwindmodel}.

\subsection{Comparison with our reference stellar models}
\label{sec:comparison}

We investigate the cases of $\logz=-2, -5$, and $-8$. The other
metallicities can be seen in Appendix~\ref{sec:OtherMetallicities}.

\begin{figure*}
  \includegraphics[width=2\columnwidth,bb=0 0 333 116]{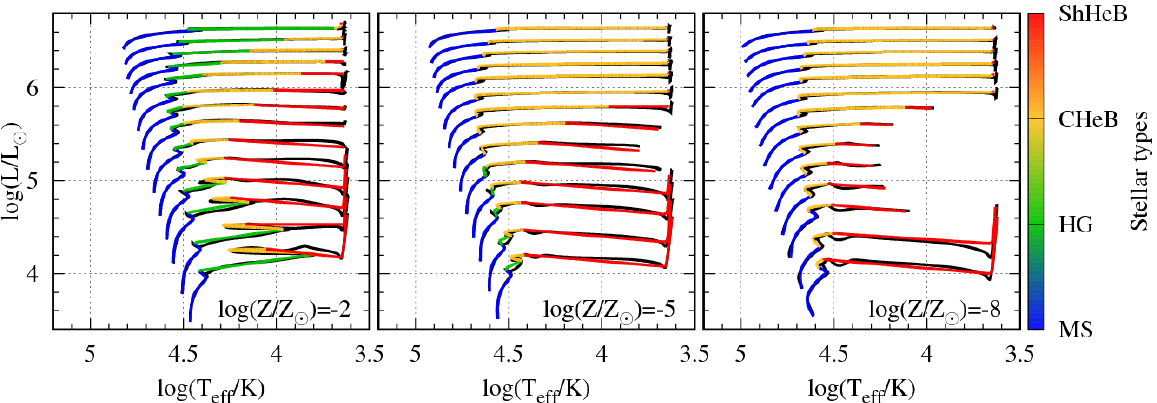}
  \caption{HR diagram for comparison between stellar models shown in
    Figure~\ref{fig:hrd} (black curves), and fitting formulae
    calculated in the \sse~code (colored curves). The color is coded
    according to stellar phases: MS, HG, CHeB, and ShHeB phases.
    \label{fig:comparison}}
\end{figure*}

We follow the time evolution of stars with $M=8, 10, 13, 16, 20, 25,
32, 40, 50, 65, 80, 100$, $125$, and $160M_\odot$ for $\logz = -2, -5$
and $-8$. In Figure~\ref{fig:comparison}, we compare our fitting
formulae with our stellar models shown in Figure~\ref{fig:hrd}. We can
see that our fitting formulae are in a good agreement with our stellar
models. For $\logz = -8$, stars with $13 \le M/M_\odot < 50$ end with
BSG stars, and other stars become RSG stars at the ending time of
their evolutions. Stars with $M/M_\odot < 13$ have entered into their
ShHeB phases by the time they become RSG stars. On the other hand,
stars with $M/M_\odot \ge 50$ still remain CHeB stars when they become
RSG stars.  For $\logz = -5$, the mass range of stars ending with BSG
stars is decreased. The mass range is $20 \le M/M_\odot <
50$. Moreover, stars with $M/M_\odot<25$ experience HG phases. Note
that no stars experience HG phases for $\logz = -8$.  We compare our
fitting formulae with our stellar models for $\logz = -2$. All the
stars enter into HG phases after MS phases. For $M/M_\odot<32$, stars
experience blue loops after the HG phases. Finally, all the stars
become RSG stars before they finish their lives. In our stellar
models, the luminosity of the star with $M=65M_\odot$ is instantly
decreased just before the star becomes a RSG star. We ignore this
instant decrease of the luminosity to make the fitting formulae.

\begin{figure*}
  \includegraphics[width=1.9\columnwidth,bb=0 0 382 226]{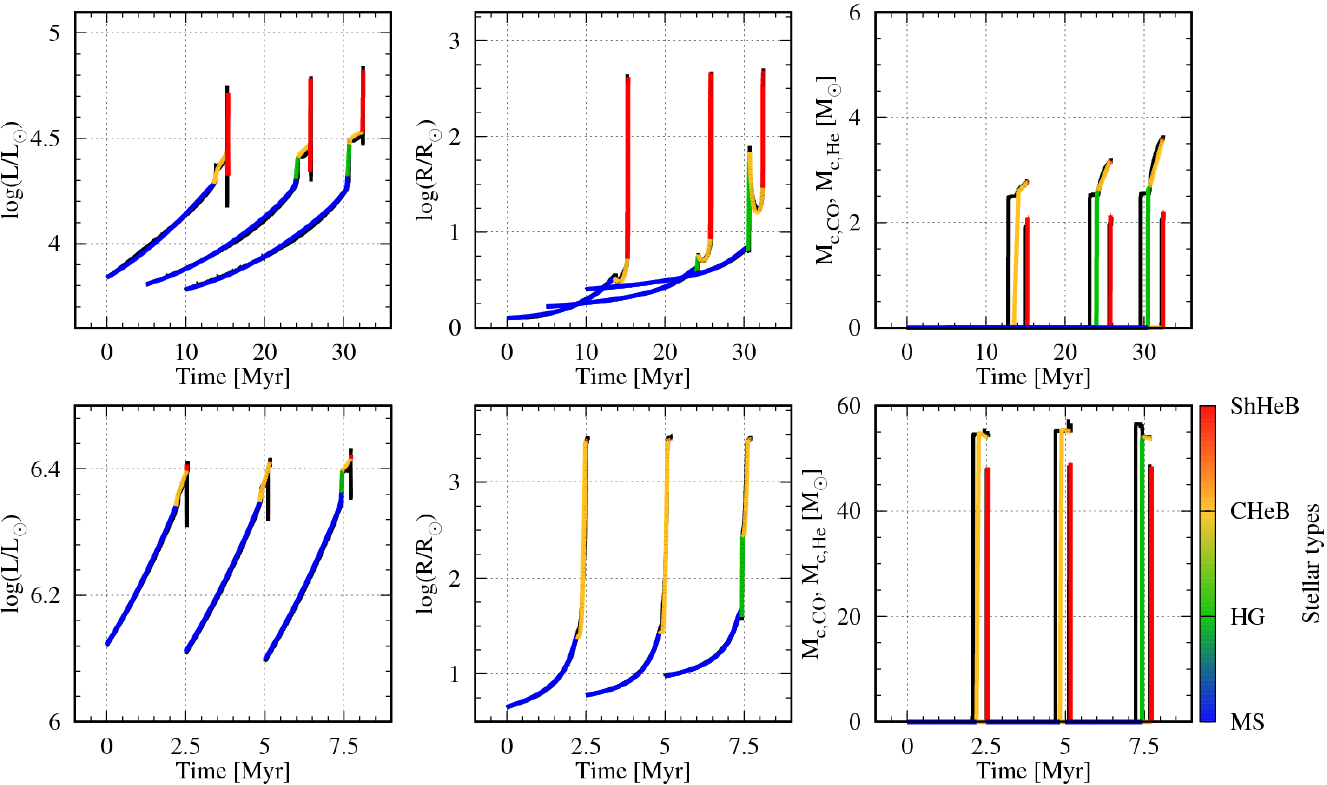}
  \caption{Time evolutions of luminosity, radius, He core mass, and CO
    core mass of stars with $M=10M_\odot$ (the top panels) and
    $M=100M_\odot$ (the bottom panels). In the rightmost panels, both
    of He and CO core masses are drawn, and the He core masses
    correspond to the larger curves. Each panel indicates these
    evolutions for $\logz=-8, -5, -2$. The times are shifted by
    $5$~Myr and $10$~Myr for $M=10M_\odot$ with $\logz = -5$ and $-2$,
    respectively, and by $2.5$~Myr and $5$~Myr for $M=100M_\odot$ with
    $\logz = -5$ and $-2$, respectively. Black solid curves indicate
    our simulation data, and colored solid curves indicate our fitting
    formulae. The color codes are the same as
    Figure~\ref{fig:comparison}. \label{fig:quantityTime}}
\end{figure*}

Figure~\ref{fig:quantityTime} shows the time evolution of stars with
$M=10$ and $100M_\odot$ for $\logz=-2, -5$, and $-8$, and compare our
fitting formulae (colored curves) with our simulation results (black
curves). The leftmost panels draw the luminosity evolutions. We can
see our fitting formulae capture features of luminosity evolutions
except for instant decrease during the CHeB or ShHeB phases in our
simulation data. The instant decrease occurs at the ending time of the
BSG phases (at temperature of $\teff \sim 10^{3.65}$~K) in our
simulation data, as seen in Figures~\ref{fig:hrd} and
\ref{fig:comparison}. Our fitting formulae deviate from our simulation
data at the instant decreases, since we ignore the instant decreases
for fitting formulae. At the ending time, our fitting formulae deviate
from our simulation data by $\Delta \log L \lesssim 0.1$. The middle
panels of Figure~\ref{fig:quantityTime} compare the radius evolutions
of our fitting formulae with those of our simulation data. These
evolutions appear in a good agreement with each other for all the
cases. The rightmost panels of Figure~\ref{fig:quantityTime} indicate
the time evolution of He and CO core masses. He cores in our fitting
formulae grow later than those in our simulation data. This is because
we set He core masses to be zero before the HG phases. CO cores in our
fitting formulae also increase later than those in our simulation
data. We assume that the CO core mass is zero before the ShHeB
phases. We quantitatively investigate these deviations and their
effects later. Note that we do not see clear metallicity dependence of
the He core mass at the end of the main-sequence. This is partly
because the overshoot effect erases metallicity dependence of He core
mass \citep{2018ApJS..237...13L}. Metallicity dependence of the He
core mass is also discussed in \citet{1986MNRAS.220..529T}.

\begin{figure}
  \includegraphics[width=0.9\columnwidth,bb=0 0 148 305]{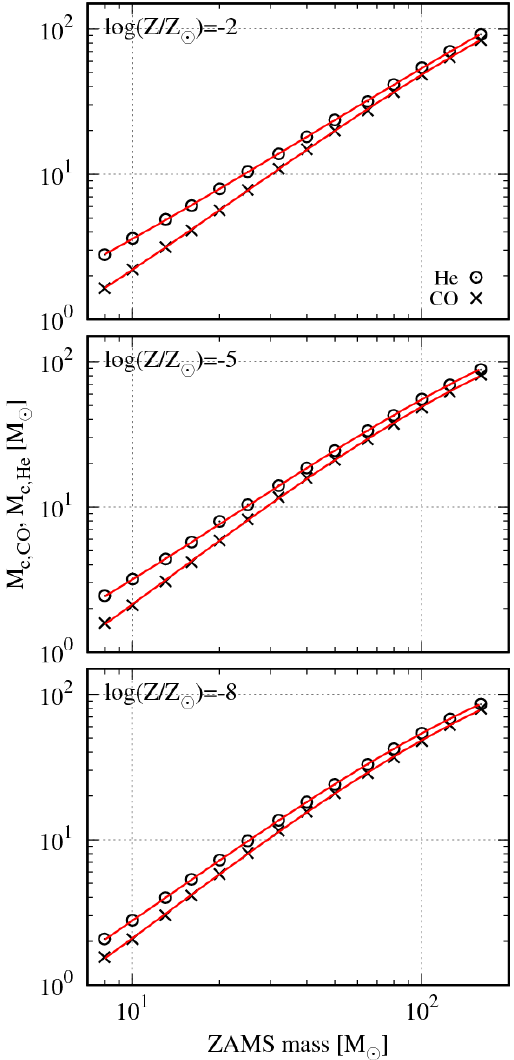}
  \caption{He (open circles) and CO (crosses) core masses at the
    ending times of the stellar evolutions as a function of ZAMS
    masses. Black points indicate our simulation data and red points
    indicate our fitting formulae. Note that the He and CO core masses
    in our fitting formulae correspond to $M_{\rm c,ShHeB}$
    (Eq.~(\ref{eq:HecoreShHeB})) and $M_{\rm c,CO}$
    (Eq.~(\ref{eq:COcoreShHeB})),
    respectively. \label{fig:coreMassFinal}}
\end{figure}

In Figure~\ref{fig:coreMassFinal}, we compare He and CO core masses in
our fitting formulae with those in our simulation data. These core
masses are $M_{\rm c,ShHeB}$ and $M_{\rm c,CO}$, i.e. ones at the
ending times of the evolutions (see Eqs.~(\ref{eq:HecoreShHeB}) and
(\ref{eq:COcoreShHeB}), respectively). They are important, since they
determine the remnant masses directly (see Eqs.(\ref{eq:Remnant}),
(\ref{eq:FeCore}), and (\ref{eq:RemnantPI})). Our fitting formulae
quite agree with our simulation data over $8 \le M/M_\odot \le
160$. We will quantitatively discuss about this in more detail below.

\begin{figure}
  \includegraphics[width=0.9\columnwidth,bb=0 0 166 116]{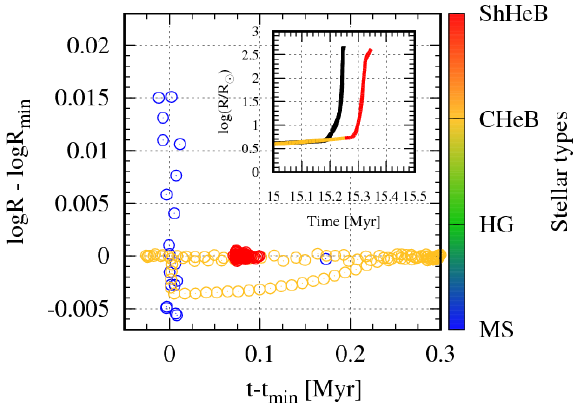}
  \caption{Deviations of radii between our fitting formulae and
    simulation data for $M=10M_\odot$ with $\logz=-8$. The deviations
    are defined in the main text. The inset figure zooms in on the
    radius evolution of the ShHeB phase. The color codes are the same
    as Figure~\ref{fig:comparison}. \label{fig:minimumErrorSample}}
\end{figure}

In order to evaluate deviations between our fitting formulae and
simulation data, we define a distance between points of our fitting
formulae and simulation data on time evolution diagrams like
Figure~\ref{fig:quantityTime} as
\begin{align}
  \Delta l = \min_j \left| \log \left( \frac{Qt}{Q_jt_j} \right)
  \right|, \label{eq:minimumError}
\end{align}
where $Q$ ($Q_j$) indicates luminosity, radius, He core mass, or CO
core mass of our fitting formulae (our simulation data) at time $t$
($t_j$). This distance is helpful to quantify not only deviations of
$Q$ between our fitting formulae and simulation data, but also
deviations of stellar evolutionary times between them. If we rescale
stellar evolutionary times of our fitting formulae so that their
evolutionary times match corresponding stellar evolutionary times of
our stellar evolution model, we should overlook the deviations of
stellar evolutionary times. If we define the deviations such that
$(Q-Q_j)/Q_j$ at time $t$, the deviations should be overestimated for
the following reason. Luminosity and radius rapidly grow at post-MS
phases. Then, slight deviations of the beginning times of some phases
(HG, CHeB, or ShHeB phases) raise large deviations of luminosities and
radii between our fitting formulae and simulation data. This can be
seen in the inset of Figure~\ref{fig:minimumErrorSample}. The
beginning time of the ShHeB phase deviates only by $\sim 0.1$~Myr,
however $(Q-Q_j)/Q_j \gtrsim 10^2$ at the beginning time.

Additionally, we define $Q_{\min}$ and $t_{\min}$ as those which are
$Q_j$ and $t_j$, respectively, taking $\Delta l$ in
Eq.~(\ref{eq:minimumError}). In other words, we can express $Q_{\min}$
and $t_{\min}$ as
\begin{align}
  \Delta l = \left| \log \left( \frac{Qt}{Q_{\min} t_{\min}} \right)
  \right|.
\end{align}
In Figure~\ref{fig:minimumErrorSample}, we show the evolution of
$(\log R - \log R_{\min}, t-t_{\min})$ for the radius evolution seen
in the inset. We can see that the radii themselves deviates by $\sim
3$~\% in the MS phase, while the time deviates by at most $0.3$~Myr in
the post-MS phases in the radius evolution.

\begin{figure*}
  \includegraphics[width=2\columnwidth,bb=0 0 357 217]{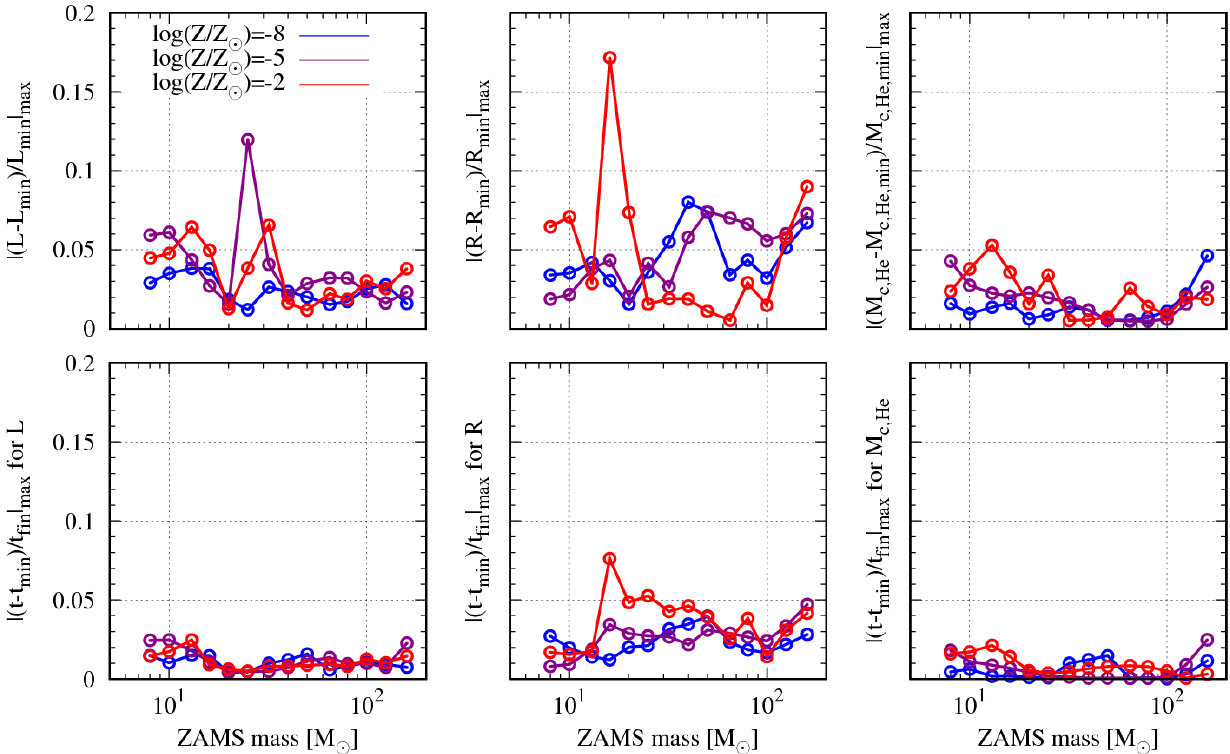}
  \caption{Maximum values of $(Q-Q_{\min})/Q_{\min}$ (top panels) and
    $(t-t_{\min})/t_{\min}$ (bottom panels) over the evolution of each
    star as a function of ZAMS mass. \label{fig:minimumErrorAll}}
\end{figure*}

Figure~\ref{fig:minimumErrorAll} shows the maximum values of
$(Q-Q_{\min})/Q_{\min}$ and $(t-t_{\min})/t_{\min}$ over the evolution
of each star. Except for the radii of $M=16M_\odot$ with $\logz=-2$,
the deviations are less than $\sim 10$~\%. Even for $M=16M_\odot$ with
$\logz=-2$, the maximum deviation of its radius is $\sim 15$~\%. It
achieves this deviation at the ending time of the HG phase.

The maximum value of $(Q-Q_{\min})/Q_{\min}$ do not fully quantify
deviations between our simulation data and fitting formulae. For
example, the luminosity at the ending time in our fitting formula is
less than that in our simulation data by $\Delta \log L \sim 0.05$ (or
by $\sim 12$~\%) for the case of $M=10M_\odot$ with $\logz=-8$ (see
the top left panel of Figure~\ref{fig:quantityTime}). On the other
hand, $\left| (L-L_{\min})/L_{\min} \right|_{\max} \sim 3$~\% as seen
in Figure~\ref{fig:minimumErrorAll}. This can be explained by the
following reason. When we choose $L_{\min}$ for the luminosity at the
ending time using Eq.~(\ref{eq:minimumError}), $L_{\min}$ is not the
luminosity at the ending time in our simulation data, but the
luminosity closest to luminosity at the ending time in our fitting
formulae.

\begin{figure*}
  \includegraphics[width=2\columnwidth,bb=0 0 352 116]{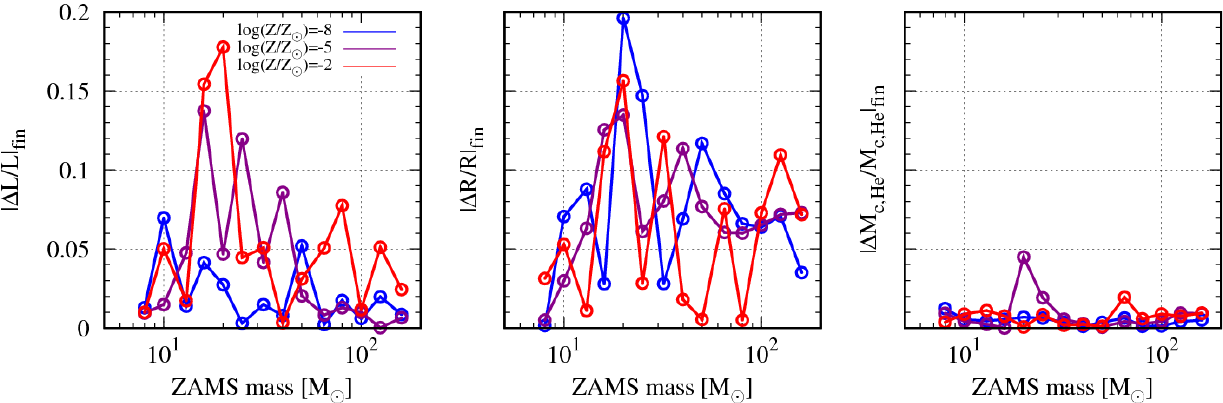}
  \caption{Deviations of quantities between our fitting formulae and
    simulation data at the ending times. \label{fig:maximumErrorAll}}
\end{figure*}

In order to solve this problem, we directly compare luminosities,
radii, and He core masses in our fitting formula with those in our
simulation data.  Figure~\ref{fig:maximumErrorAll} shows these
quantities at the ending time for our fitting formulae and simulation
data. Then, we find these deviations are $\sim 20$~\% at most, which
is the radius at the ending time of $M=20M_\odot$ with
$\logz=-8$. These deviations in luminosities and radii tend to be
large near $M/M_\odot \sim 20$ for the following reason. Stars with
$M/M_\odot \lesssim 20$ reach to RSG stars, while stars with
$M/M_\odot \gtrsim 20$ end with BSG stars. The luminosities and radii
at the ending times strongly depend on stellar masses. Simple
polynomials we use for our fitting formulae cannot follow such strong
dependence. In contrast with the luminosities and radii at the ending
time, the He and CO core masses at the ending time in our fitting
formulae are different from in our simulation data by at most $5$~\%.

Eventually, our fitting formulae deviate from our simulation data by
at most $20$~\%. We can say our fitting formulae have deviations small
enough to be used for population synthesis calculations and star
cluster simulations. These deviations are much smaller than
uncertainties contained in population synthesis calculations and star
cluster simulations, such as common envelope evolution.

\subsection{Combination with a stellar wind model}
\label{sec:combination}

In this section, we couple our fitting formulae with a stellar wind
model described in section~\ref{sec:stellarwindmodel}, and investigate
stellar evolutions and their remnants.

\begin{figure*}
  \includegraphics[width=1.9\columnwidth,bb=0 0 347 213]{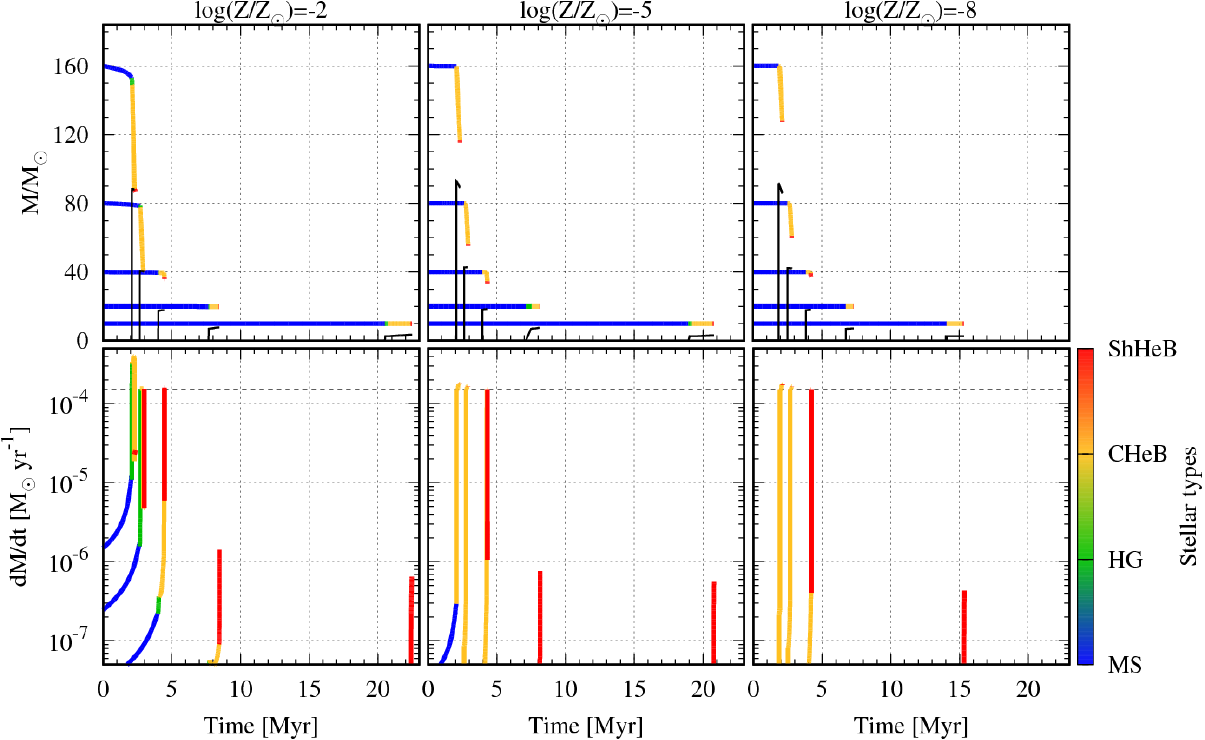}
  \caption{Time evolution of the total mass (top) and stellar wind
    mass loss (bottom) of stars with $10, 20, 40, 80$, and
    $160M_\odot$ for $\logz=-2, -5$, and $-8$. We can distinguish
    which curve is which stellar mass by the ending time of the
    evolution; the ending time is earlier as the stellar mass is
    larger. Black solid curves in the top panels indicate the time
    evolution of He core masses.  Black dashed lines in the bottom
    panels indicate the mass loss of LBV stars $\dot{M}_{\rm LBV}=1.5
    \times 10^{-4} M_\odot\mbox{yr}^{-1}$. The color codes are the
    same as Figure~\ref{fig:comparison}. \label{fig:windEvolution}}
\end{figure*}

Figure~\ref{fig:windEvolution} shows time evolution of the total mass
and stellar wind mass loss of stars with different masses and
metallicities. Stars with $M/M_\odot \lesssim 40M_\odot$ receive
little stellar wind mass loss for all the metallicities. This is owing
to low metallicities. Stars with $M/M_\odot \gtrsim 80M_\odot$
significantly lose their masses in their CHeB phases. This mass loss
is driven by LBV winds, $\dot{M}_{\rm LBV}$. This is true for all the
metallicities, since the LBV mass loss does not depend on metallicity
in our model. They lose their masses in their CHeB phases rather than
in their ShHeB phases. This is mainly because the durations of their
CHeB phases are longer than those of their ShHeB phases. The other
reason is that a part of stars become naked He stars described below.

Stars with $\logz = -2$ and $M/M_\odot \gtrsim 80M_\odot$ additionally
receive the mass loss of luminous stars, $\dot{M}_{\rm NJ}$, from
their MS phases to their CHeB phases. Note that their highest mass
loss rates exceed the LBV mass loss rate $\dot{M}_{\rm LBV}=1.5 \times
10^{-4}M_\odot\mbox{yr}^{-1}$. As a result of this, they become naked
He stars. We can confirm this from the fact that their total masses
are equal to their He core masses.  When they are naked He stars,
their mass loss rates are $\sim 6 \times 10^{-6}
M_\odot\mbox{yr}^{-1}$ for $M/M_\odot=80$, and $\sim 3 \times 10^{-5}
M_\odot\mbox{yr}^{-1}$ for $M/M_\odot=160$, dominated by the mass loss
of WR stars. They might be observed as WR stars.

\begin{figure}
  \includegraphics[width=0.9\columnwidth,bb=0 0 145 116]{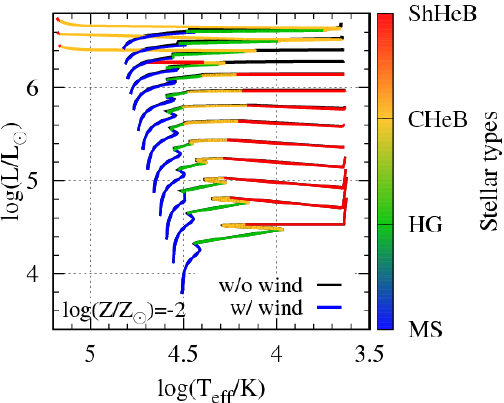}
  \caption{HR diagram of stars for $\logz = -2$ with and without
    stellar wind mass loss (red and black curves, respectively). We do
    not show the evolution of a star with $M=8M_\odot$ at the ZAMS
    time, since its mass is decreased to $M/M_\odot<8$ due to the mass
    loss, and deviates from the scope of application of our fitting
    formulae. The color codes are the same as
    Figure~\ref{fig:comparison}. \label{fig:windEffect}}
\end{figure}

We can follow stellar evolution to a naked He star by our fitting
formulae, taking into account stellar wind mass loss by a
post-processing way. Figure~\ref{fig:windEffect} shows the HR diagram
of stars for $\logz = -2$ with and without stellar wind mass loss. The
evolutions of stars with $M/M_\odot \le 65$ are similar regardless of
the presence and absence of the mass loss. On the other hand, stars
with $M/M_\odot \ge 80$ evolve blueward after the He ignition. This is
because they become naked He stars for $M/M_\odot > 80$, and nearly a
naked He star for $M/M_\odot = 80$ due to the mass loss. The mass
boundary might be different from other evolution tracks due to
difference of stellar evolution and wind models. However, the
important point is that such a post-processing way to include stellar
wind mass loss can represent the evolution to a naked He star.

\begin{figure}
  \includegraphics[width=0.9\columnwidth,bb=0 0 173 169]{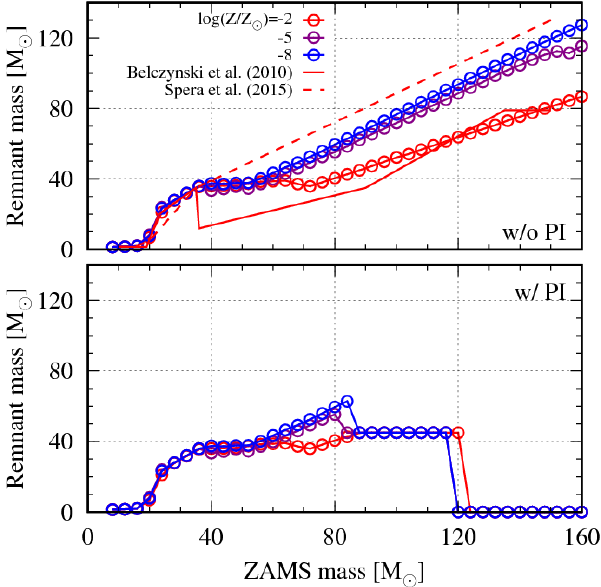}
  \caption{Relation between ZAMS star and their remnant masses for the
    w/o PI and w/ PI models (top and bottom, respectively). In the top
    panel, we add the relations of Belczynski's and Spera's models for
    $\logz = -2$
    \citep[][respectively]{2010ApJ...714.1217B,2015MNRAS.451.4086S}. We
    read these data by eye from the
    literature. \label{fig:remnantMass}}
\end{figure}

Figure~\ref{fig:remnantMass} shows the relation between ZAMS and
remnant masses. We first focus on the w/o PI model. We compare our
$\logz=-2$ results with those of \cite{2010ApJ...714.1217B}. Stars
with $M/M_\odot \lesssim 20$ leave NSs, which is consistent with
them. BH masses left by stars with $20 \lesssim M/M_\odot \lesssim 35$
are also in good agreement with them.  BH masses left by stars with
$M/M_\odot \gtrsim 60$ are larger than theirs by $10M_\odot$, since
our He and CO core masses are more massive by
$10M_\odot$. Nevertheless, the trend of the remnant masses in our
fitting formulae is in good agreement with that in
\cite{2010ApJ...714.1217B}. There is large discrepancy in BH masses in
the range of stellar masses $35 \lesssim M/M_\odot \lesssim 60$. These
stars in our model receive LBV winds on shorter durations, since they
have radii large enough to satisfy the LBV criterion (see
Eq.~(\ref{eq:dotMLBV})) on shorter durations. We find BH masses left
by these stars are different among different evolution
tracks. $40M_\odot$ stars leave BHs with $39M_\odot$ in our model,
$15M_\odot$ in \cite{2010ApJ...714.1217B}, and $36M_\odot$ in
\texttt{SEVN} \citep[see fig.~14 in][]{2015MNRAS.451.4086S}. We
conclude this discrepancy does not matter. The remnant masses for
$\logz = -5$, $-8$ are larger than for $\logz = -2$, since stellar
wind mass loss becomes weaker with metallicity decreasing. However,
the remnant masses are similar between the cases of $\logz = -5$ and
$-8$. Stellar wind mass loss becomes ineffective for EMP stars, and
does not sensitively depend on metallicity in this regime.

In $M/M_\odot \lesssim 80$, remnant masses in the w/ PI model are the
same as in the w/o PI model. Otherwise, the former are smaller than
the latter for all the metallicities due to PPI and PI~SN effects.
Remnant masses are $45M_\odot$ in $80 \lesssim M/M_\odot \lesssim 120$
for all the metallicities. PPI work in this mass range. This is
consistent with the relation between ZAMS and He core masses in
Figure~\ref{fig:coreMassFinal} which shows $45 \lesssim M_{\rm
  c,He}/M_\odot \lesssim 65$ in $80 \lesssim M/M_\odot \lesssim
120$. In $M/M_\odot \gtrsim 120$, remnant masses become zero, since
these stars experience PI~SNe. For $\logz=-5$ and $-8$, stars with $M
\sim 80M_\odot$ achieve the maximum remnant masses exceeding
$45M_\odot$. This is because they do not undergo PPI nor PI~SNe due to
small He core masses, and leave hydrogen envelope owing to their low
metallicities. On the other hand, the maximum remnant mass is
$45M_\odot$ for $\logz = -2$. Although stars with $M \sim 65M_\odot$
do not experience PPI nor PI~SNe, they have no hydrogen envelope, and
evolve to naked He stars at the ending time (see
Figure~\ref{fig:windEffect}). The lower mass limits of stars
undergoing PPI and PI~SNe in our fitting formulae are smaller than in
the model of \cite{2016A&A...594A..97B}. This is because He core
masses in our fitting formula are larger than in their model, as
described above.

\begin{figure}
  \includegraphics[width=0.9\columnwidth,bb=0 0 173 96]{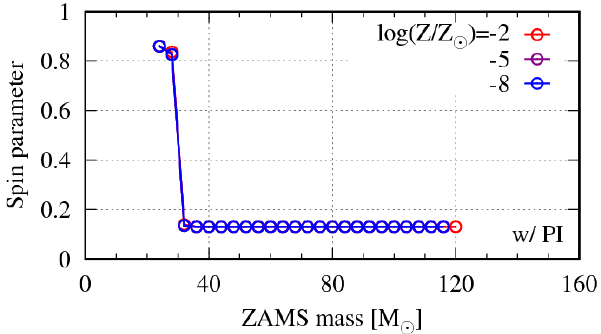}
  \caption{Relation between ZAMS star and their remnant spins for the
    w/ PI models. \label{fig:remnantSpin}}
\end{figure}

Figure~\ref{fig:remnantSpin} shows remnant spins as a function of ZAMS
mass for the w/ PI model. There is no data point in the mass range of
$M/M_\odot \lesssim 20$ or $M/M_\odot \gtrsim 120$. This is because
stars with $M/M_\odot \lesssim 20$ leave NSs, and those with
$M/M_\odot \gtrsim 120$ leave no remnant due to PI~SNe. BHs for the
w/o PI model have the same spins when $M \lesssim 120M_\odot$. For the
w/o PI, stars with $M \gtrsim 120M_\odot$ leave BHs with spins as low
as stars with $M \sim 120M_\odot$ do.

\section{Summary}
\label{sec:summary}

We have devised the fitting formulae of EMP stars. Their metallicities
are $\logz = -2, -4, -5, -6$, and $-8$. The fitting formulae consider
stars ending with BSG stars, and stars skipping HG phases and blue
loops. In our fitting formulae, relatively light stars still remain
BSG stars when they finish their CHeB phases. On the other hand, all
the stars finish their BSG phases before they finish their CHeB phases
in the Hurley's models. This is not true especially for relatively
light stars. Therefore, our modeling can be more realistic than the
Hurley's models. Our fitting formulae are in good agreement with our
stellar models, which are consistent with Marigo's model. Our fitting
formulae can be used on the \sse, \bse, \nbodyfour, and
\nbodysix\;codes for population synthesis calculations and star
cluster simulations. We believe they should be useful to elucidate the
origin of merging BH-BHs observed by gravitational wave observatories.

\section*{Acknowledgments}

We thank Hurley~J.~R. and Wang~L. for making \bse\;and
\texttt{NBODY6++GPU} open sources, respectively. This research has
been supported in part by Grants-in-Aid for Scientific Research
(16K17656, 17K05380, 17H01130, 17H06360, 18J00558,
19K03907) from the Japan Society for the Promotion of Science.

\appendix

\section{Values for fitting formula}
\label{sec:valuesForFittingFormula}

We show constants used in section~\ref{sec:implementation}.

\begin{table*}
  \centering
  \caption{Constants for $\logz = -2$.}
  \label{tab:massLimit-2}
  \begin{tabular}{lllll}
     $i$ & 0 & 1 & 2 & 3 \\
     \hline
    ${\cal T}_{{\rm HeI},i}$       & +1.5839255604016800e+00 & +7.9580658743321393e+01 & +6.2605049204196496e+02 & +4.9099525171608302e+03 \\
    ${\cal T}_{{\rm CHeB},i}$      & +1.9589322976358600e-01 & +7.8483404642259504e+00 & +4.5085546669210199e+00 & +6.5327035939216103e+02 \\
    ${\cal T}_{{\rm EBSG},i}$      & +1.5491443477301301e+00 & +1.0157506729497200e+02 & +4.1045797870126700e+02 & +6.5666106502774601e+03 \\
    ${\cal T}_{{\rm Fin},i}$       & +1.7829748399131200e+00 & +8.7662108101897800e+01 & +6.2980590516040002e+02 & +5.6078890004744198e+03 \\
    ${\cal L}_{{\rm ZAMS},i}$      & -4.3387121453610497e-02 & +4.7002611105589196e+00 & -9.3228362169805801e-01 & +5.8811945206418803e-02 \\
    ${\cal L}_{{\rm EMS},i}$       & -1.4163054579500700e-01 & +6.2468925492728600e+00 & -2.0517336248547200e+00 & +2.7740223008074300e-01 \\
    ${\cal L}_{{\rm HeI},i}$       & +9.1569000891833996e-01 & +4.5656469439486997e+00 & -1.1002164667539900e+00 & +9.3840594819720097e-02 \\
    ${\cal L}_{{\rm ECHeB},i}$     & +7.0984159927752299e-01 & +5.2269835165592902e+00 & -1.6253670199780199e+00 & +2.1924040178420401e-01 \\
    ${\cal L}_{{\rm EBSG},i}$      & +1.6658494286226599e+00 & +2.8135842651513001e+00 & +1.0831049212944700e-01 & -1.6438606141683301e-01 \\
    ${\cal L}_{{\rm Fin,l},i}$     & +7.1935042363085397e+00 & -1.0757949651891300e+01 & +1.2454591128743299e+01 & -4.0650097608005700e+00 \\
    ${\cal L}_{{\rm Fin,u},i}$     & -1.0863563795046700e+00 & +7.9383414627732600e+00 & -2.9992532262892801e+00 & +4.5364789567682801e-01 \\
    ${\cal L}_{{\rm \alpha},i}$    & -1.9408020240768799e-02 & -8.9435341439184898e-02 & +4.5147634586394197e-01 & -1.5123156474279600e-01 \\
    ${\cal L}_{{\rm \beta},i}$     & -1.6671325813774601e-01 & +4.3653653543205401e-01 & -3.1149021748090500e-01 & +6.7838974435627503e-02 \\
    ${\cal L}_{{\rm \Delta},i}$    & +6.7428636579205697e-02 & +3.7982090032742701e-02 & -5.7959714605753401e-02 & +1.3089024896359399e-02 \\
    ${\cal R}_{{\rm ZAMS},i}$      & -3.6870298415699199e-01 & +9.4401642726269397e-01 & -2.1017534522620901e-01 & +3.7277105006281598e-02 \\
    ${\cal R}_{{\rm EMS},i}$       & -1.4359719418587300e+00 & +4.1335756478450900e+00 & -2.4940809801056401e+00 & +5.9090072146249595e-01 \\
    ${\cal R}_{{\rm HeI},i}$       & +4.9292595459292903e+00 & -3.2253557317799100e+00 & -7.3958869377935499e-01 & +8.6851132826960398e-01 \\
    ${\cal R}_{{\rm min},i}$       & -3.0780806973814201e+00 & +1.0260779580627400e+01 & -8.2059416870071100e+00 & +2.2289738933720402e+00 \\
    ${\cal R}_{{\rm CHeB,EBSG},i}$ & +6.1925984270592203e+00 & -9.6076904024682896e+00 & +5.6240568009227401e+00 & -7.3097573061153098e-01 \\
    ${\cal R}_{{\rm Fin},i}$       & +0.0000000000000000e+00 & +0.0000000000000000e+00 & +0.0000000000000000e+00 & +0.0000000000000000e+00 \\
    ${\cal R}_{{\rm \alpha},i}$    & -7.3471934070371600e-02 & +2.5091437411072098e-01 & -4.1162900537159798e-02 & +8.9919312090114400e-03 \\
    ${\cal R}_{{\rm \beta},i}$     & -3.8906641158399602e-01 & +1.1515583275404400e+00 & -8.7408225106827897e-01 & +2.7666584341436701e-01 \\
    ${\cal R}_{{\rm \gamma},i}$    & -1.0404715460267899e+00 & +2.3176110463747399e+00 & -1.7072675527739301e+00 & +3.8380480274952700e-01 \\
    ${\cal R}_{{\rm \Delta},i}$    & -2.0879641699855300e-01 & +3.7006200560629499e-01 & -2.5730658470289097e-01 & +4.8661473352442100e-02 \\
    ${\cal R}_{{\rm RSG},0i}$      & -8.7074249061240602e-02 & -1.6922018399772801e-01 \\
    ${\cal R}_{{\rm RSG},1i}$      & +6.1896196949691396e-01 & -5.7427447948830397e-03 \\
    ${\cal H}_{{\rm HeI},i}$       & -1.2316548227442199e+00 & +1.8848911925803100e+00 & -2.4609484339633700e-01 & +2.2463422369182098e-02 \\
    ${\cal H}_{{\rm ECHeB},i}$     & -3.8303715910210101e-01 & +7.0150741599662503e-01 & +2.9607706988601801e-01 & -5.9315517041934301e-02 \\
    ${\cal C}_{{\rm CO},i}$        & -7.3806577478145996e-01 & +7.0263944811078005e-01 & +5.0208830062187804e-01 & -1.2376219079451400e-01 \\
    \hline
  \end{tabular}
\end{table*}

\begin{table*}
  \centering
  \caption{Constants for $\logz = -4$.}
  \label{tab:massLimit-4}
  \begin{tabular}{lllll}
     $i$ & 0 & 1 & 2 & 3 \\
     \hline
    ${\cal T}_{{\rm HeI},i}$       & +1.5517508063147400e+00 & +7.8762231265283901e+01 & +6.6306742996156697e+02 & +3.9780303119567002e+03 \\
    ${\cal T}_{{\rm CHeB},i}$      & +2.5995643378195399e-01 & +1.3758291938603699e+00 & +1.1749452241249701e+02 & +7.3578237381483504e+01 \\
    ${\cal T}_{{\rm EBSG},i}$      & +1.5746855412155101e+00 & +9.3014443586965996e+01 & +5.9278989148185997e+02 & +4.8922266924294800e+03 \\
    ${\cal T}_{{\rm Fin},i}$       & +1.8136363987309900e+00 & +8.0465190932245406e+01 & +7.7654726489562199e+02 & +4.1226848652793597e+03 \\
    ${\cal L}_{{\rm ZAMS},i}$      & +7.0523683128112099e-03 & +4.6392179589624503e+00 & -9.0127242744985803e-01 & +5.3238721546632901e-02 \\
    ${\cal L}_{{\rm EMS},i}$       & -9.5090922675450204e-02 & +6.1861769329705298e+00 & -2.0227268480957301e+00 & +2.7233787294898498e-01 \\
    ${\cal L}_{{\rm HeI},i}$       & +5.5593067547365704e-01 & +5.2576512230847499e+00 & -1.5733536006708599e+00 & +1.9982532305519199e-01 \\
    ${\cal L}_{{\rm ECHeB},i}$     & +1.0036121148808901e+00 & +4.4477153532156501e+00 & -1.0517047979711500e+00 & +9.0669105159825694e-02 \\
    ${\cal L}_{{\rm EBSG},i}$      & +1.9138600333723099e+00 & +2.0401605565817502e+00 & +6.8545215127409798e-01 & -2.9136581751254698e-01 \\
    ${\cal L}_{{\rm Fin,l},i}$     & +1.3665205743983201e+01 & -2.8581615325979701e+01 & +2.8565817023847401e+01 & -8.8588379270868707e+00 \\
    ${\cal L}_{{\rm Fin,u},i}$     & -1.5295137595677000e+01 & +3.1558544210358502e+01 & -1.5855688265929601e+01 & +2.7497620458708600e+00 \\
    ${\cal L}_{{\rm \alpha},i}$    & +7.7337632457581199e-02 & -2.4681139438928901e-01 & +5.3387040906391603e-01 & -1.6557459200983801e-01 \\
    ${\cal L}_{{\rm \beta},i}$     & -4.3175574474877097e-02 & +1.7703088876154599e-01 & -1.3612814349929700e-01 & +3.0317560049377499e-02 \\
    ${\cal L}_{{\rm \Delta},i}$    & -2.6732530743815101e-02 & +2.2559320802858901e-01 & -1.7984675493080099e-01 & +3.8317535205556301e-02 \\
    ${\cal R}_{{\rm ZAMS},i}$      & -4.3620639472716299e-01 & +8.4978687662735197e-01 & -1.5993111424465600e-01 & +2.7669651740859301e-02 \\
    ${\cal R}_{{\rm EMS},i}$       & -1.0109813992939201e+00 & +2.8983622792063399e+00 & -1.6345733256974599e+00 & +4.0423625972161897e-01 \\
    ${\cal R}_{{\rm HeI},i}$       & +1.7618554313938800e-01 & +2.0515198024096599e+00 & -1.8411912436347599e+00 & +6.0716387191588606e-01 \\
    ${\cal R}_{{\rm min},i}$       & -1.1462751987327799e+00 & +4.0914478967912702e+00 & -2.8790729602077900e+00 & +7.8151962602801806e-01 \\
    ${\cal R}_{{\rm CHeB,EBSG},i}$ & +1.6237359045461699e+00 & +5.6527800027967900e-01 & -3.1779617931783601e+00 & +2.0224444898627301e+00 \\
    ${\cal R}_{{\rm Fin},i}$       & +0.0000000000000000e+00 & +0.0000000000000000e+00 & +0.0000000000000000e+00 & +0.0000000000000000e+00 \\
    ${\cal R}_{{\rm \alpha},i}$    & -3.0660053188372399e-01 & +7.3518486490284396e-01 & -3.7303325190354403e-01 & +8.0660868408699404e-02 \\
    ${\cal R}_{{\rm \beta},i}$     & -1.1764079419992901e+00 & +2.8256595551415500e+00 & -2.0106042556923298e+00 & +5.1511742066562305e-01 \\
    ${\cal R}_{{\rm \gamma},i}$    & +2.6666916897183601e-02 & -5.6086994492167697e-02 & -3.6193625945596802e-02 & +1.4815537338410900e-02 \\
    ${\cal R}_{{\rm \Delta},i}$    & -1.7247613923205901e-01 & +3.1501048172396900e-01 & -2.3143156092273801e-01 & +4.9808789329703798e-02 \\
    ${\cal R}_{{\rm RSG},0i}$      & +9.0940251890292804e-03 & -2.3234236135069800e-01 \\
    ${\cal R}_{{\rm RSG},1i}$      & +5.9863657659087000e-01 & +6.6527853775347704e-03 \\
    ${\cal H}_{{\rm HeI},i}$       & -1.0535947893773301e+00 & +1.4170998603813501e+00 & +1.2689928777198800e-01 & -6.7494753025814203e-02 \\
    ${\cal H}_{{\rm ECHeB},i}$     & -3.9425104148142198e-01 & +4.9002611835324500e-01 & +5.7239586075853399e-01 & -1.4093689615634999e-01 \\
    ${\cal C}_{{\rm CO},i}$        & -9.4823722226647700e-01 & +9.7236386558568799e-01 & +4.5283017654016999e-01 & -1.3877489041026900e-01 \\
    \hline
  \end{tabular}
\end{table*}

\begin{table*}
  \centering
  \caption{Constants for $\logz = -5$.}
  \label{tab:massLimit-5}
  \begin{tabular}{lllll}
     $i$ & 0 & 1 & 2 & 3 \\
     \hline
    ${\cal T}_{{\rm HeI},i}$       & +1.5608149013984800e+00 & +7.3567160614509504e+01 & +8.1835557597302898e+02 & +2.1085273006198800e+03 \\
    ${\cal T}_{{\rm CHeB},i}$      & +2.5190328005023799e-01 & +2.3937574545125400e+00 & +6.8538719197550407e+01 & +3.4090149355802299e+02 \\
    ${\cal T}_{{\rm EBSG},i}$      & +1.6461473568952700e+00 & +8.5024735569825793e+01 & +7.5746222377079403e+02 & +3.0364037858196398e+03 \\
    ${\cal T}_{{\rm Fin},i}$       & +1.8149674275343299e+00 & +7.6275812519002400e+01 & +8.8345559090139704e+02 & +2.5200002713951699e+03 \\
    ${\cal L}_{{\rm ZAMS},i}$      & +6.1565122879347303e-02 & +4.5519847009521301e+00 & -8.4909794442826203e-01 & +4.2747189247281603e-02 \\
    ${\cal L}_{{\rm EMS},i}$       & -6.4010588830543502e-02 & +6.1613002916478603e+00 & -2.0261739188010601e+00 & +2.7603556655566103e-01 \\
    ${\cal L}_{{\rm HeI},i}$       & +5.8122492862095898e-01 & +5.1451646393334300e+00 & -1.4989396222973901e+00 & +1.8578769559084299e-01 \\
    ${\cal L}_{{\rm ECHeB},i}$     & +8.0634329880297295e-01 & +4.7633074342534902e+00 & -1.2175693546732000e+00 & +1.1848267301948801e-01 \\
    ${\cal L}_{{\rm EBSG},i}$      & +1.4195127851226099e+00 & +2.9963262421921000e+00 & +9.7803246316281900e-02 & -1.7525207469906401e-01 \\
    ${\cal L}_{{\rm Fin,l},i}$     & +1.4494199842344100e+01 & -3.2273248268857600e+01 & +3.3509908142265203e+01 & -1.0945672335926499e+01 \\
    ${\cal L}_{{\rm Fin,u},i}$     & -2.9400093428201099e+00 & +1.0683813612299200e+01 & -4.2900521722222003e+00 & +6.4299798091772997e-01 \\
    ${\cal L}_{{\rm \alpha},i}$    & +1.3507674415184301e-01 & -3.1004292468722500e-01 & +5.5482963687162401e-01 & -1.6843671132819599e-01 \\
    ${\cal L}_{{\rm \beta},i}$     & -1.5512993576365000e-01 & +3.7628702204329800e-01 & -2.4439907829143701e-01 & +4.8734830791039803e-02 \\
    ${\cal L}_{{\rm \Delta},i}$    & +6.8922515715162704e-03 & +1.7942852148415400e-01 & -1.6879676205781699e-01 & +3.9630241617149800e-02 \\
    ${\cal R}_{{\rm ZAMS},i}$      & -4.2008197743730302e-01 & +7.1601089423309805e-01 & -8.7939534847640000e-02 & +1.4674567587019200e-02 \\
    ${\cal R}_{{\rm EMS},i}$       & -9.2456828602432894e-01 & +2.5389191763683798e+00 & -1.3751410532902100e+00 & +3.5045831707709002e-01 \\
    ${\cal R}_{{\rm HeI},i}$       & +3.9738040817917603e-01 & +6.3392838342428004e-01 & -4.8249880551521102e-01 & +2.1465700473096400e-01 \\
    ${\cal R}_{{\rm min},i}$       & -6.4199204960415601e-01 & +2.4849515765003098e+00 & -1.5602241696515400e+00 & +4.2012270812299601e-01 \\
    ${\cal R}_{{\rm CHeB,EBSG},i}$ & -7.8927435380812403e+00 & +2.2851238614423899e+01 & -2.0357967863018199e+01 & +6.3389279382339296e+00 \\
    ${\cal R}_{{\rm Fin},i}$       & +2.4064199986630800e+02 & -4.8538695379477701e+02 & +3.2781854087263798e+02 & -7.3251789698885403e+01 \\
    ${\cal R}_{{\rm \alpha},i}$    & -3.1420648674110302e-01 & +7.3720759135681202e-01 & -3.6998786143878298e-01 & +7.8226901219150699e-02 \\
    ${\cal R}_{{\rm \beta},i}$     & -1.1538548374112800e+00 & +2.7792891314058101e+00 & -2.0035406799845701e+00 & +5.2103603281852995e-01 \\
    ${\cal R}_{{\rm \gamma},i}$    & -9.9705770420161696e-03 & +3.4151857848480502e-02 & -1.2150727797915800e-01 & +4.6426009888991697e-02 \\
    ${\cal R}_{{\rm \Delta},i}$    & -1.2237289829584599e-01 & +1.7993028395168600e-01 & -1.1080605018453900e-01 & +1.9014426091513000e-02 \\
    ${\cal R}_{{\rm RSG},0i}$      & +5.6659069526797702e-02 & -2.9401839565042698e-01 \\
    ${\cal R}_{{\rm RSG},1i}$      & +5.9229869792013501e-01 & +1.5810282790358201e-02 \\
    ${\cal H}_{{\rm HeI},i}$       & -1.1011433924550000e+00 & +1.5272663127372501e+00 & +3.5517282030031301e-02 & -4.3877721910985303e-02 \\
    ${\cal H}_{{\rm ECHeB},i}$     & -4.1085136748565099e-01 & +4.4292411042459900e-01 & +6.2270607618818596e-01 & -1.5313454615055200e-01 \\
    ${\cal C}_{{\rm CO},i}$        & -9.2895961573096897e-01 & +9.2331358192538004e-01 & +4.7945677789672903e-01 & -1.4307351573737101e-01 \\
    \hline
  \end{tabular}
\end{table*}

\begin{table*}
  \centering
  \caption{Constants for $\logz = -6$.}
  \label{tab:massLimit-6}
  \begin{tabular}{lllll}
     $i$ & 0 & 1 & 2 & 3 \\
     \hline
    ${\cal T}_{{\rm HeI},i}$       & +1.5056408142507200e+00 & +7.4994995850899201e+01 & +8.3116022167663198e+02 & +4.9040743535153001e+02 \\
    ${\cal T}_{{\rm CHeB},i}$      & +3.0667170921629699e-01 & -2.2573597515177801e+00 & +1.3262687600114401e+02 & -4.3276720558498297e+01 \\
    ${\cal T}_{{\rm EBSG},i}$      & +1.6655342649798499e+00 & +8.1652799673668696e+01 & +8.2906385668603400e+02 & +1.0852909614256901e+03 \\
    ${\cal T}_{{\rm Fin},i}$       & +1.8146885357058899e+00 & +7.3040756282165205e+01 & +9.6032575420450996e+02 & +5.2267066025411305e+02 \\
    ${\cal L}_{{\rm ZAMS},i}$      & +8.6457066168576402e-02 & +4.5267539598251902e+00 & -8.3734602904147404e-01 & +4.0772686211378001e-02 \\
    ${\cal L}_{{\rm EMS},i}$       & +7.0757975602064205e-02 & +5.9547056937739100e+00 & -1.9224648832190800e+00 & +2.5857233194367901e-01 \\
    ${\cal L}_{{\rm HeI},i}$       & +4.4142301118520499e-01 & +5.2539507603229101e+00 & -1.4936383116696199e+00 & +1.7333956725116800e-01 \\
    ${\cal L}_{{\rm ECHeB},i}$     & +5.5163281110466300e-01 & +5.2726483070966701e+00 & -1.5669073962725100e+00 & +1.9653692229092701e-01 \\
    ${\cal L}_{{\rm EBSG},i}$      & +2.3653114492122501e-01 & +5.3880332036350200e+00 & -1.4180348147378099e+00 & +1.3120380751216901e-01 \\
    ${\cal L}_{{\rm Fin,l},i}$     & -7.8189143203214698e+01 & +2.3910265122146399e+02 & -2.2922323244477499e+02 & +7.3069269426524997e+01 \\
    ${\cal L}_{{\rm Fin,u},i}$     & +1.1685971748214700e-01 & +5.9355760414335199e+00 & -1.9145262506356799e+00 & +2.5840425254712202e-01 \\
    ${\cal L}_{{\rm \alpha},i}$    & +7.1130274997986598e-01 & -1.3554969336771401e+00 & +1.1690764274913901e+00 & -2.8733405856569799e-01 \\
    ${\cal L}_{{\rm \beta},i}$     & +8.7864061232266005e-02 & -8.2310583604875406e-02 & +2.2756522018622102e-02 & -1.2808221435716401e-03 \\
    ${\cal L}_{{\rm \Delta},i}$    & +2.8862524329672300e-02 & +1.2503925155034501e-01 & -1.2931759799313100e-01 & +3.0882782606178400e-02 \\
    ${\cal R}_{{\rm ZAMS},i}$      & -3.3237368336527201e-01 & +4.6306428086272700e-01 & +4.7820892172886598e-02 & -9.6166327966181395e-03 \\
    ${\cal R}_{{\rm EMS},i}$       & -4.3397491671777300e-01 & +1.3377594599674700e+00 & -5.1421774804168696e-01 & +1.5412807397524700e-01 \\
    ${\cal R}_{{\rm HeI},i}$       & +6.1860120021200900e-02 & +4.0036628051026202e-01 & +5.9419501733342803e-02 & +4.0113003159862798e-02 \\
    ${\cal R}_{{\rm min},i}$       & -1.4216357295539300e-01 & +7.8608012916942305e-01 & -1.7661790431570801e-01 & +8.7027362522545407e-02 \\
    ${\cal R}_{{\rm CHeB,EBSG},i}$ & -3.6832391211242701e+00 & +1.1084061918623000e+01 & -9.7324743496085198e+00 & +3.1373519906345702e+00 \\
    ${\cal R}_{{\rm Fin},i}$       & +5.7804594901412303e+00 & +4.2481043663905496e+00 & -1.2931309320913600e+01 & +5.6039641623863998e+00 \\
    ${\cal R}_{{\rm \alpha},i}$    & -3.8233593929888698e-01 & +7.9002072854240402e-01 & -3.4866555880093603e-01 & +5.9835915587617299e-02 \\
    ${\cal R}_{{\rm \beta},i}$     & -7.1524193355817201e-01 & +1.4559437723950699e+00 & -8.4329698201959102e-01 & +2.2410177619832200e-01 \\
    ${\cal R}_{{\rm \gamma},i}$    & +3.4301337795213799e-02 & -1.2266252485565400e-01 & +8.4527293107284907e-02 & -2.7605337667557499e-02 \\
    ${\cal R}_{{\rm \Delta},i}$    & +5.0499661728335798e-01 & -1.0261689627690800e+00 & +6.2043697099474304e-01 & -1.2325834401596100e-01 \\
    ${\cal R}_{{\rm RSG},0i}$      & -1.1360958843644700e-01 & -1.4734021414496001e-01 \\
    ${\cal R}_{{\rm RSG},1i}$      & +6.2134621256327505e-01 & -7.9113517758866905e-03 \\
    ${\cal H}_{{\rm HeI},i}$       & -1.1406439008233300e+00 & +1.5754309620831299e+00 & +1.5150861500077399e-02 & -4.1230152708572498e-02 \\
    ${\cal H}_{{\rm ECHeB},i}$     & -5.2437528299791603e-01 & +5.5932229149602297e-01 & +5.7797948042505098e-01 & -1.4589837221339999e-01 \\
    ${\cal C}_{{\rm CO},i}$        & -9.2137012965923404e-01 & +8.8636637162070298e-01 & +5.0592667147908199e-01 & -1.4855704392558100e-01 \\
    \hline
  \end{tabular}
\end{table*}

\begin{table*}
  \centering
  \caption{Constants for $\logz = -8$.}
  \label{tab:massLimit-8}
  \begin{tabular}{lllll}
     $i$ & 0 & 1 & 2 & 3 \\
     \hline
    ${\cal T}_{{\rm HeI},i}$       & +1.2305856460495901e+00 & +1.0134578934187600e+02 & +9.1468769124419097e+01 & +1.8034769714629199e+03 \\
    ${\cal T}_{{\rm CHeB},i}$      & +2.4927377844451301e-01 & +2.1430677547749402e+00 & +8.5972144678894594e+01 & -1.4028396903502400e+02 \\
    ${\cal T}_{{\rm EBSG},i}$      & +1.3730451469334499e+00 & +1.0966949368516001e+02 & +8.9197507907499698e+01 & +2.0930245791822999e+03 \\
    ${\cal T}_{{\rm Fin},i}$       & +1.4810249724314000e+00 & +1.0389495324882300e+02 & +1.7328483361378301e+02 & +1.7439035999222599e+03 \\
    ${\cal L}_{{\rm ZAMS},i}$      & -4.9125370974699000e-03 & +4.7854346156440304e+00 & -1.0218914733694000e+00 & +8.0491993338893300e-02 \\
    ${\cal L}_{{\rm EMS},i}$       & +1.8683454969478600e-01 & +5.7145070422311797e+00 & -1.7657692128941300e+00 & +2.2589703362508501e-01 \\
    ${\cal L}_{{\rm HeI},i}$       & +1.8683454969478600e-01 & +5.7145070422311797e+00 & -1.7657692128941300e+00 & +2.2589703362508501e-01 \\
    ${\cal L}_{{\rm ECHeB},i}$     & +3.8790608169726498e-01 & +5.4894852836696097e+00 & -1.6450220435261300e+00 & +2.0133843339575599e-01 \\
    ${\cal L}_{{\rm EBSG},i}$      & -7.8446922484086201e-01 & +7.4136655401824401e+00 & -2.6966654623657602e+00 & +3.9086840832154501e-01 \\
    ${\cal L}_{{\rm Fin,l},i}$     & -5.2063766360667403e+00 & +1.9423911281843100e+01 & -9.4740006457763393e+00 & +0.0000000000000000e+00 \\
    ${\cal L}_{{\rm Fin,u},i}$     & +2.2104250108866999e-01 & +5.4053774623698096e+00 & -1.3814930395748199e+00 & +1.1265166679862200e-01 \\
    ${\cal L}_{{\rm \alpha},i}$    & +8.4548309048483195e-01 & -1.2539875520793100e+00 & +9.2320801185925605e-01 & -2.0896705174094601e-01 \\
    ${\cal L}_{{\rm \beta},i}$     & -3.4823782284743299e-01 & +6.6175971466374905e-01 & -4.0197845268117299e-01 & +7.9552888490381807e-02 \\
    ${\cal L}_{{\rm \Delta},i}$    & +1.6333532063424699e-01 & -1.4798579565892800e-01 & +4.7496374698078603e-02 & -5.8983127069122503e-03 \\
    ${\cal R}_{{\rm ZAMS},i}$      & +4.6071669041284602e-01 & -1.2482170192596500e+00 & +1.1079694078714399e+00 & -2.1778915454158501e-01 \\
    ${\cal R}_{{\rm EMS},i}$       & -8.7660234566099804e-01 & +1.9194682681297499e+00 & -7.3392245022216696e-01 & +1.7274738264264999e-01 \\
    ${\cal R}_{{\rm HeI},i}$       & -8.7660234566099804e-01 & +1.9194682681297499e+00 & -7.3392245022216696e-01 & +1.7274738264264999e-01 \\
    ${\cal R}_{{\rm min},i}$       & -8.7660234566099804e-01 & +1.9194682681297499e+00 & -7.3392245022216696e-01 & +1.7274738264264999e-01 \\
    ${\cal R}_{{\rm CHeB,EBSG},i}$ & -1.8343478577029000e+01 & +4.7783881233948101e+01 & -4.0169462871126797e+01 & +1.1460630304015901e+01 \\
    ${\cal R}_{{\rm Fin},i}$       & -7.8795250912748704e+00 & +2.8961085855700901e+01 & -2.8088036390052000e+01 & +8.7716640685062703e+00 \\
    ${\cal R}_{{\rm \alpha},i}$    & -2.8687484575077300e+00 & +5.4361579138011802e+00 & -3.0890032437297399e+00 & +6.0545880904935201e-01 \\
    ${\cal R}_{{\rm \beta},i}$     & -2.0581094227135899e+00 & +3.8644269698339500e+00 & -2.2704304648914801e+00 & +5.0131612800208902e-01 \\
    ${\cal R}_{{\rm \gamma},i}$    & +6.2100077845129797e-01 & -1.2611619209677001e+00 & +7.8193448082669703e-01 & -1.6755954354525801e-01 \\
    ${\cal R}_{{\rm \Delta},i}$    & +9.9983829393963303e-02 & -4.4410431106289799e-01 & +3.5725587872207698e-01 & -8.6199515162683704e-02 \\
    ${\cal R}_{{\rm RSG},0i}$      & +1.9892249955808999e-01 & -3.7299109886527798e-01 \\
    ${\cal R}_{{\rm RSG},1i}$      & +5.6071811071005395e-01 & +3.2722143200640798e-02 \\
    ${\cal H}_{{\rm HeI},i}$       & -1.0733886890363700e+00 & +1.4584928984676999e+00 & +8.1519136854884905e-02 & -5.3730041751472800e-02 \\
    ${\cal H}_{{\rm ECHeB},i}$     & -7.2712749189720804e-01 & +8.3176655071820205e-01 & +4.7853206104410101e-01 & -1.3980539673247300e-01 \\
    ${\cal C}_{{\rm CO},i}$        & -9.5225083520154796e-01 & +9.4357015904211505e-01 & +4.7429337580493902e-01 & -1.4351236491880501e-01 \\
    \hline
  \end{tabular}
\end{table*}

\section{Other metallicities}
\label{sec:OtherMetallicities}

\begin{figure}
  \includegraphics[width=0.9\columnwidth,bb=0 0 146 207]{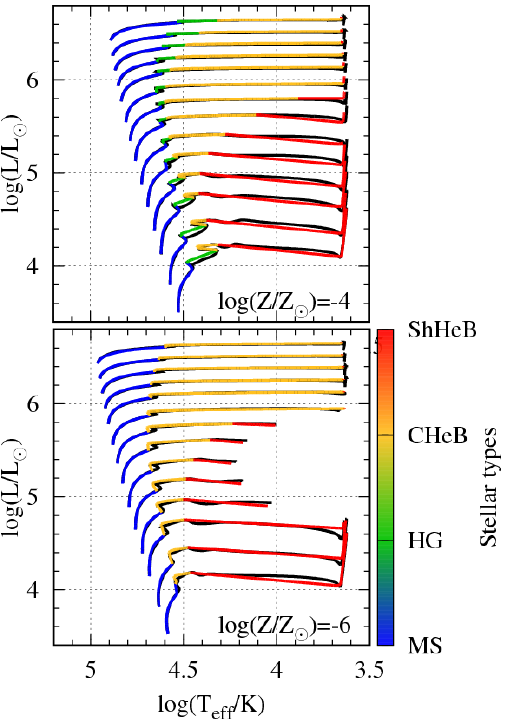}
  \caption{The same as Figure~\ref{fig:comparison}, except for $\logz
    = -4$ and $-6$. \label{fig:comparisonAppendix}}
\end{figure}

\begin{figure*}
  \includegraphics[width=1.9\columnwidth,bb=0 0 382 226]{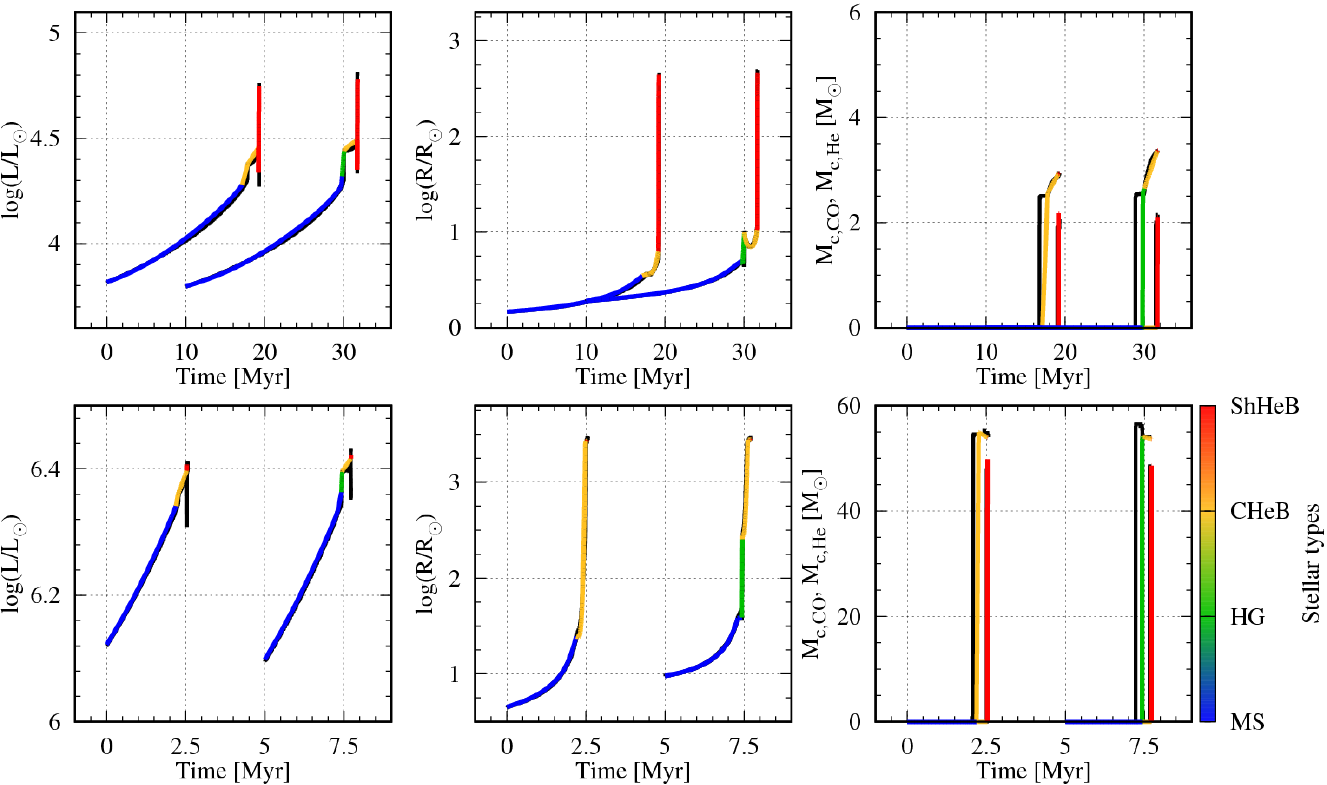}
  \caption{The same as Figure~\ref{fig:quantityTimeAppendix}, except
    for $\logz = -4$ and $-6$. The times are shifted by $10$~Myr for
    $M=10M_\odot$ with $\logz = -4$, and by $5$~Myr for $M=100M_\odot$
    with $\logz = -2$. \label{fig:quantityTimeAppendix}}
\end{figure*}

\begin{figure}
  \includegraphics[width=0.9\columnwidth,bb=0 0 136
    209]{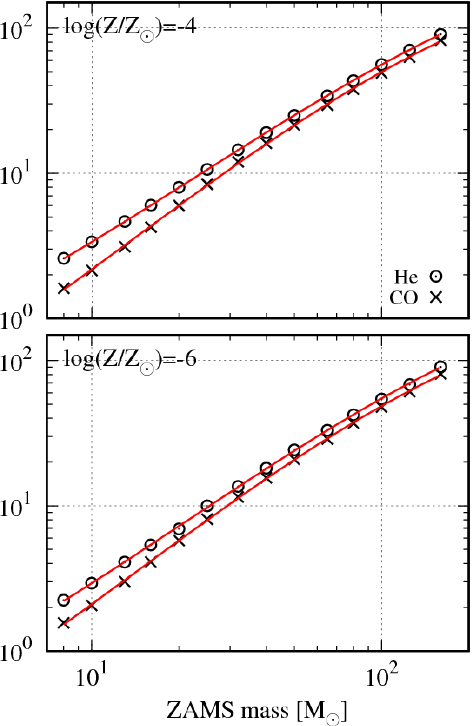}
  \caption{The same as Figure~\ref{fig:coreMassFinal}, except for
    $\logz = -4$ and $-6$. \label{fig:coreMassFinalAppendix}}
\end{figure}

In this section, we present comparison between our fitting formulae
and reference stellar models for $\logz = -4$ and
$-6$. Figures~\ref{fig:comparisonAppendix},
\ref{fig:quantityTimeAppendix}, and \ref{fig:coreMassFinalAppendix}
corresponds to Figure~\ref{fig:comparison}, \ref{fig:quantityTime},
and \ref{fig:coreMassFinal}, respectively. They match well each other.

\bibliographystyle{mnras}

\bsp	
\label{lastpage}
\end{document}